\documentclass[preprint2,numberedappendix]{emulateapj-rtx4}
\usepackage{graphicx,natbib}
\usepackage{bm,url,color}
\graphicspath{{./fig/}{./png/}}

\newcommand{\EQ}{\begin{equation}}
\newcommand{\EN}{\end{equation}}
\newcommand{\EQA}{\begin{eqnarray}}
\newcommand{\ENA}{\end{eqnarray}}

\newcommand{\Eq}[1]{Equation~(\ref{#1})}
\newcommand{\Eqs}[2]{Equations~(\ref{#1}) and~(\ref{#2})}

\newcommand{\Sec}[1]{Sect.~\ref{#1}}

\newcommand{\Fig}[1]{Figure~\ref{#1}}
\newcommand{\FFig}[1]{Figure~\ref{#1}}

\newcommand{\Figs}[2]{Figures~\ref{#1} and \ref{#2}}

\newcommand{\Tab}[1]{Table~\ref{#1}}
\newcommand{\Tabs}[2]{Tables~\ref{#1} and \ref{#2}}

\newcommand{\bra}[1]{\langle #1\rangle}

{}
{}
{}

{}
{}
{}
{}
{}
{}
{}
{}
{}
{}
{}
{}
{}
{}
{}
{}
{}

{}

{}

{}
{}
{}

\newcommand{\zzz}{\hat{\mbox{\boldmath $z$}} {}}

\newcommand{\nullvector}{{\bf0}}

\newcommand{\kk}{\bm{k}}

\newcommand{\xx}{\bm{x}}

\newcommand{\BB}{\bm{B}}

\newcommand{\JJ}{\bm{J}}

\newcommand{\AAA}{\bm{A}}

\newcommand{\ee}{\bm{e}}

\newcommand{\ff}{\mbox{\boldmath $f$} {}}

\newcommand{\nab}{{\bm{\nabla}}}

\newcommand{\ii}{{\rm i}}

\newcommand{\const}{{\rm const}  {}}

\def\Rey{\mbox{\rm Re}}

\def\Brms{B_{\rm rms}}

\def\half{{\textstyle{1\over2}}}

\newcommand{\T}{\,{\rm T}}
\newcommand{\G}{\,{\rm G}}

\newcommand{\K}{\,{\rm K}}

\newcommand{\s}{\,{\rm s}}

\newcommand{\cm}{\,{\rm cm}}

\newcommand{\m}{\,{\rm m}}
\newcommand{\km}{\,{\rm km}}

\newcommand{\Myr}{\,{\rm Myr}}

\newcommand{\erg}{\,{\rm erg}}

\newcommand{\J}{\,{\rm J}}

\newcommand{\A}{\,{\rm A}}

\newcommand{\yjpp}[3]{ #1, {JPlPh,} {#2}, #3}
\newcommand{\yapj}[3]{ #1, {ApJ,} {#2}, #3}

\newcommand{\yana}[3]{ #1, {A\&A,} {#2}, #3}

\newcommand{\ypf}[3]{ #1, {PhFl,} {#2}, #3}

\newcommand{\ypp}[3]{ #1, {PhPl,} {#2}, #3}

\newcommand{\yprl}[3]{ #1, {PhRvL,} {#2}, #3}

\newcommand{\yptrsa}[3]{ #1, {RSPTA,} {#2}, #3}
\newcommand{\ymn}[3]{ #1, {MNRAS,} {#2}, #3}

\newcommand{\yprd}[3]{ #1, {PhRvD,} {#2}, #3}
\newcommand{\ypre}[3]{ #1, {PhRvE,} {#2}, #3}
\newcommand{\ypnas}[3]{ #1, {PNAS,} {#2}, #3}

\newcommand{\yjour}[4]{ #1, {#2}, {#3}, #4}

\begin{document}

\title{Hall cascade with fractional magnetic helicity in neutron star crusts}

\author{
Axel Brandenburg$^{1,2,3}$\thanks{E-mail:brandenb@nordita.org}
}

\affil{
$^1$Nordita, KTH Royal Institute of Technology and Stockholm University, Roslagstullsbacken 23, SE-10691 Stockholm, Sweden\\
$^2$Department of Astronomy, AlbaNova University Center, Stockholm University, SE-10691 Stockholm, Sweden\\
$^3$McWilliams Center for Cosmology \& Department of Physics, Carnegie Mellon University, Pittsburgh, PA 15213, USA
}

\submitted{Astrophys. J., accepted (2020)}
\date{Received 2020 June 22; revised 2020 August 3; accepted 2020 August 3; $ $Revision: 1.107 $ $ $\!$}

\begin{abstract}
The ohmic decay of magnetic fields in the crusts of neutron stars is
generally believed to be governed by Hall drift which leads to what
is known as a Hall cascade.
Here we show that helical and fractionally helical magnetic fields
undergo strong inverse cascading like in magnetohydrodynamics (MHD),
but the magnetic energy decays more slowly with time $t$:
$\propto t^{-2/5}$ instead of $\propto t^{-2/3}$ in MHD.
Even for a nonhelical magnetic field there is a certain degree of
inverse cascading for sufficiently strong magnetic fields.
The inertial range scaling with wavenumber $k$ is compatible with earlier
findings for the forced Hall cascade, i.e., proportional to $k^{-7/3}$,
but in the decaying cases, the subinertial range spectrum steepens to a
novel $k^5$ slope instead of the $k^4$ slope in MHD.
The energy of the large-scale magnetic field can increase quadratically
in time through inverse cascading.
For helical fields, the energy dissipation is found to be inversely
proportional to the large-scale magnetic field and proportional to the
fifth power of the root-mean square (rms) magnetic field.
For neutron star conditions with an rms magnetic field of a few times
$10^{14}\G$, the large-scale magnetic field might only be $10^{11}\G$,
while still producing magnetic dissipation of $10^{33}\erg\s^{-1}$ for
thousands of years, which could manifest itself through X-ray emission.
Finally, it is shown that the conclusions from local unstratified models
agree rather well with those from stratified models with boundaries.
\end{abstract}


\section{Introduction}

Over the first hundreds of years after the freezing of the crust of a
neutron star (NS), magnetic dissipation is believed to power the X-ray
emission observed in the central compact objects of supernova remnants.
At the same time, the large-scale magnetic field, as characterized by
its dipole field strength, is not strong enough to explain this directly
as a result of magnetic dissipation; see \cite{GWH16,GWH18,GHI20} for
the motivation.
Moreover, the magnetic field would decay too slowly to explain the
observed emission.
A plausible mechanism may therefore be the ``turbulent'' decay of a
small-scale magnetic field in the NS's crust \citep{VCO00}.
Such an enhanced decay with correspondingly enhanced Joule dissipation
could be driven by the nonlinearity from the Hall effect \citep{HR02,HR04}.
Following \cite{GR92}, we refer to this process simply as Hall cascade,
keeping in mind that no motions are involved.

Traditionally, NS magnetic fields are explained as the result of
compressive amplification of a large-scale magnetic field in the
NS's progenitor.
However, this explanation ignores the fact that the NS is fully convective
during the first tens of seconds of its lifetime \citep{Eps79}.
Not only would this have destroyed a preexisting magnetic field, but
it would have produced a potentially much stronger one from scratch
\citep{TD93}.
Compared with the short time scales of such NS convection, the rotation
is usually slow.
It is therefore questionable whether this process alone could explain
the large-scale magnetic field in supernova remnants.
Instead, it is possible that the crust of the NS is dominated by a
small-scale (turbulent) magnetic field at the time of freezing.
However, rotation may still be responsible for causing the turbulence
to be at least partially helical.
This could be crucial for moderating the speed of the decay of the
small-scale magnetic field.
It could also explain the gradual amplification of a large-scale
magnetic field through inverse cascading \citep{Cho11}.

In the NS crust, ions are immobile, so the electric current $\JJ$
is carried by the electrons alone.
Their velocity is therefore $-\JJ/en_e$, where $e$ is the
elementary charge and $n_e$ is the number density of electrons
\citep[see, e.g.,][]{CL09}.
The evolution of the magnetic field $\BB$ with time $t$ is then given by
\EQ
\frac{\partial\BB}{\partial t}=
\nab\times\left(-\frac{\JJ\times\BB}{en_e}-\eta\mu_0\JJ\right),\quad
\JJ=\frac{1}{\mu_0}\nab\times\BB,\;
\label{dBdt}
\EN
where $\eta=1/\mu_0\sigma_{\rm el}$ is the magnetic diffusivity with
$\mu_0$ being the magnetic permeability, and $\sigma_{\rm el}$ is the
electric conductivity.
The nonlinearity in \Eq{dBdt} resembles that of the vorticity equation
in hydrodynamics \citep{GR92}, although there are also significant
differences.
For example, \cite{WH10} noted that, unlike usual turbulence, where
smaller eddies are advected by larger ones, this is not the case in
the Hall cascade.
It is also known that the inertial range follows a $k^{-7/3}$ magnetic
energy spectrum with wavenumber $k$ \citep{Bis+96,Bis+99}, which is
steeper than the Kolmogorov $k^{-5/3}$ kinetic energy spectrum.

Meanwhile, significant progress has been made in understanding the
decay of hydrodynamic and magnetohydrodynamics (MHD) turbulence,
both with and without magnetic helicity.
The work of \cite{BK17} used the instantaneous scaling exponents $p$
and $q$ in the scalings of mean energy density ${\cal E}\propto t^{-p}$
and correlation length $\xi\propto t^q$ versus time $t$.
They found that, as the solution approached selfsimilar scaling,
$p$ and $q$ settled toward a specific point in a $pq$ diagram.
Particularly familiar cases are $p=10/7$ and $q=2/7$ when the Loitsiansky
integral is conserved \citep{BP56}, $p=6/5$ and $q=2/5$ when the Saffman
integral is conserved \citep{Saf67}, or $p=q=2/3$ when the magnetic
helicity is conserved \citep{BM99}.
In view of these new diagnostics, it is timely to revisit the evolution
of magnetic fields in the Hall cascade.

In this paper, we study the initial value problem of \Eq{dBdt} for
small-scale (turbulent) magnetic fields.
Unlike \cite{RG02}, who considered an initially large-scale magnetic field
that becomes unstable and then leads to the production of small-scales,
we follow here the proposal of \cite{GR92} and consider the case of
an initially small-scale field that continues to decay, but with the
possibility of a nonvanishing magnetic helicity, which is a conserved
quantity also in the Hall cascade \citep{Cho11}.

In addition to the turbulent conversion into Joule heat, we study
the power law decay of magnetic energy and of the peak wavenumber
of the energy spectrum.
We restrict ourselves to Cartesian geometry with coordinates
$\xx=(x,y,z)$.
We focus on the case of triply periodic domains.
This facilitates the use of Fourier spectra as our principal means of
diagnostics and is best suited to address generic decay properties of
the Hall cascade.
However, to address the relevance of these idealized models to
real NS crusts, we also consider stratified cases where the electron
density and electric conductivity increase with height in an approximately
realistic way.
In those cases we use nonperiodic boundary conditions in the vertical
direction.
The use of standard three-dimensional Fourier transformation is then
still possible, but as a diagnostic means it is not ideal unless Fourier
transformation is only employed in the horizontal direction.
We begin by presenting our basic model and study some of its relevant
properties.

\section{The model}

\subsection{Units and NS parameters}
\label{UnitsNS}

We consider a Hall cascade with a characteristic wavenumber $k_0$,
which is where the spectrum peaks initially.
It is related to the spherical harmonic degree $\ell$, where most of
the energy resides, through $k_0=\ell/R$, where $R$ is the NS radius.
For reasons that will be given below, we will consider time-dependent
values of $\eta$ for many of our models.
We therefore also define a representative constant $\eta_0$ that can be
chosen to be equal to the initial value of $\eta$.
In the stratified cases, the surface values of $\eta$ and $n_e$
are denoted by $\eta_0$ and $n_{e0}$, respectively.
We present the results in nondimensional form by introducing the
following units
\EQ
[\xx]=k_0^{-1},\quad [t]=(\eta_0 k_0^2)^{-1},\quad
[\BB]=e n_{e0} \mu_0 \eta_0.
\label{Units}
\EN
This implies that the current density is measured in units of
$[\JJ]=[\BB]\,k_0/\mu_0$.
We will also be interested in the magnetic dissipation,
$\epsilon=\eta\mu_0\bra{\JJ^2}$.
It has dimensions of energy density per unit time, or
$[\epsilon]=e^2 n_{e0}^2 \mu_0 \eta_0^3 k_0^2$.

Using $\mu_0=4\pi\times10^{-7}\T\m\A^{-1}$, $e=1.60\times10^{-19}\A\s$,
$n_{e0}=2.5\times10^{40}\m^{-3}$, and $\eta_0=4\times10^{-8}\m^2\s^{-1}$
\citep{GWH16,GHI20}, we have $[B]=2\times10^{12}\G$, where $1\G=10^{-4}\T$
has been used.
With $\ell=10$ \citep{GWH16,GHI20} and $R=10^4\m$, we have
$k_0=10^{-3}\m^{-1}$, so $[t]=0.8\Myr$.
We also have $[\epsilon]=1.3\times10^{9}\J\m^{-3}\s^{-1}$.
To obtain the total (electromagnetic and neutrino) luminosity $L$
from Joule dissipation, we have to multiply $\epsilon$ by the volume,
which we take to be $10^{12}\m^3$ for a $1\km$ thick layer around the NS.
For the luminosity, we then find $[L]=1.3\times10^{28}\erg\s^{-1}$,
where we have used $1\J=10^7\erg$.
To express our simulation results in dimensionful units, we multiply by
the appropriate units given above.

Before introducing fully nondimensional units in the next section,
we point out that one could introduce a normalized magnetic field as
$\BB'=\BB/en_e\mu_0$, which has dimensions of $\m^2\s^{-1}$, i.e.,
the same as the magnetic diffusivity, and also the same as the velocity
potential (scalar and vector).
This is a natural choice, but it is somewhat unexpected given that in
MHD one rather tends to think of the magnetic field as a velocity.
This difference is significant in that it implies a dimensional argument
for the resulting turbulence spectrum that is different from that in MHD.
We return to this in \Sec{DimensionalArgument}.

\subsection{The basic equation}

In \Eq{dBdt}, we replace $\BB=\BB' e n_e\mu_0\eta_0$, to scale out both
the Hall coefficient and $\eta_0$ in \Eq{dBdt}; see, e.g., \cite{RG02}.
This implies that instead of varying the Hall coefficient, we study the
behavior for different magnetic field strengths.
For the rest of this paper, we drop the primes.

To preserve the solenoidality of the magnetic field at all times, it is
convenient to solve the induction equation with the Hall term for the
magnetic vector potential $\AAA$.
In the unstratified case, we solve the equation
\EQ
\frac{\partial\AAA}{\partial t}=-\JJ\times\BB-\eta\JJ+\ff,
\quad\JJ=\nab\times\BB,
\label{dAdt}
\EN
where $\BB=\nab\times\AAA$ is the magnetic field in terms of the magnetic
vector potential, and $\ff$ is a stochastic forcing function that is
used in some our cases studied below.
The minus sign in \Eq{dAdt} is insignificant and could have been scaled
out as well.

In the stratified case, we choose the domain to be in the range
$-d\leq z\leq0$, where $d$ is the depth and $z=0$ is the position of
the surface.
We adopt the profile function $\zeta(z)=(1-z/H_e)^4$ with $n_e\propto\zeta$ and
$\eta\propto\zeta^{-2/3}$ \citep{GWH16,GHI20}, where $H_e$ is the scale height
for the electron density and $\zeta=1$ at the top.
This formulation is accurate enough for the purpose of the present
investigation, although more realistic profiles could be computed; see
\cite{CH08} for a review on the theory of the stratification of neutron
star crusts.
Instead of \Eq{dAdt}, we now solve
\EQ
\frac{\partial\AAA}{\partial t}=-\frac{\JJ\times\BB}{\zeta}
-\frac{\eta\JJ}{\zeta^{2/3}},
\label{dAdt2}
\EN
and instead of periodic boundary conditions, we use a pseudo-vacuum
condition ($\zzz\times\BB=\nullvector$) on $z=0$ and a perfect conductor
condition ($\zzz\times\AAA=\nullvector$ with $\zzz\cdot\BB=0$) on $z=-d$.

In the unstratified, triply periodic cases, we consider two types
of initial fields: one with a broken power law spectrum and one that
is obtained by driving the system for a short amount of time with a
monochromatic forcing.
In the stratified case, we only consider a broken power law spectrum
without initial forcing, which is why we have omitted the $\ff$ term
in \Eq{dAdt2}.
These procedures are described in the following two sections.

\subsection{Broken power law initial conditions, $\ff=\nullvector$}

As in \cite{BK17} and \cite{BKMRPTV17}, we construct the initial condition
for the magnetic vector potential ${\bm A}({\bm x},0)$ from a random
three-dimensional vector field in real space that is $\delta$-correlated
in space.
In the following, hats denote Fourier transformation in all three directions.
We transform this field into Fourier space and construct the magnetic field
as $\hat{\BB}(\kk)=i\kk\times\hat{\AAA}(\kk)$.
We then scale the magnetic field by a function of $k=|\kk|$ such that we
obtain the desired initial spectrum.
We also apply the projection operator $P_{ij}=\delta_{ij}-k_i k_j/k^2$
to make $\AAA$ divergence free, and add a certain fraction $\sigma$
(not to be confused with the electric conductivity $\sigma_{\rm el}$)
to make the resulting field helical.
Thus, we have
\begin{equation}
B_i({\kk})=B_0\left(P_{ij}(\kk)-\ii\sigma\epsilon_{ijl}
k_l/k\right) g_j({\kk})\, S(k),
\label{Bikk}
\end{equation}
where $g_j(\kk)$ is the Fourier transform of a $\delta$-correlated
vector field in three dimensions with Gaussian fluctuations, $k_0$
is now identified with the initial wavenumber of the energy-carrying
eddies, and $S(k)$ determines the spectral shape with \citep{BKMRPTV17}
\begin{equation}
S(k)={k_0^{-3/2} (k/k_0)^{\alpha/2-1}
\over[1+(k/k_0)^{2(\alpha+7/3)}]^{1/4}}.
\label{Sfunction}
\end{equation}
For a given value of $B_0$, the resulting initial value of the root-mean
square (rms) magnetic field $\Brms$,
which will be denoted by $\Brms^{(0)}$, is usually somewhat larger.
For $k_0/k_1=180$, for example, we find $\Brms^{(0)}/B_0\approx3.2$
when $\sigma=0$, and $\Brms^{(0)}/B_0\approx4.5$ when $\sigma=1$.

This broken power law initial condition is also used in the stratified
cases.
The application of nonperiodic boundary conditions in the $z$ direction
may cause sharp gradients in places, but this never led to any noticeable
effects.

\subsection{Monochromatic initial driving, $\ff\neq\nullvector$}

In some cases, we apply in \Eq{dAdt} monochromatic forcing with the
$\ff$ term during a short initial time interval $0\leq t\leq t_{\rm ini}$
to produce an initial condition for the rest of the simulation, when
$\ff=\nullvector$.
In some cases, when we are interested in stationary turbulence, we also
keep $\ff\neq\nullvector$ during the entire time of the simulation.
When forcing is on, we select randomly at each time step a phase
$-\pi<\varphi\le\pi$ and the components of the wavevector $\kk$ from
many possible discrete wavevectors in a certain range around a given
value $k_0$.
In this way, the adopted forcing function
\begin{equation}
\ff(\xx,t)={\rm Re}\{{\cal N}\tilde{\ff}(\kk,t)\exp[i\kk\cdot\xx+i\varphi]\}
\end{equation}
is white noise in time and consists of plane waves with average wavenumber
$k_0$.
Here, $\xx$ is the position vector and
${\cal N}=[\JJ][\BB](\eta_0k_0^2\delta t)^{1/2}$
is a normalization factor, where $\delta t$ is the time step.
The Fourier amplitudes are
\begin{equation}
\tilde{f}_i=\left(\delta_{ij}-i\sigma\epsilon_{ijl} k_l/k\right)
\tilde{f}_j^{(0)}/\sqrt{1+\sigma^2},\;
\end{equation}
where $\tilde{\ff}^{\rm(0)}({\kk})=(\kk\times\ee)/[\kk^2-(\kk\cdot\ee)^2]^{1/2}$
is a nonhelical forcing function.
Here, $\ee$ is an arbitrary unit vector that are not aligned with $\kk$.
Note that $|\ff|^2=1$.
We consider both $\sigma=0$ and $\sigma=1$,
corresponding to nonhelical and maximally helical cases.
The forcing is only enabled during the time interval $0\leq t\leq t_1$,
where $t_1$ is the actual starting time of the simulation.
In this sense, this forcing procedure can be considered as part of the initial condition.

\subsection{Spectral diagnostics}
\label{SpectralDiagnostics}

In the triply periodic cases, we study the evolution of magnetic energy
and magnetic helicity spectra which are defined as \citep[cf.][]{BN11}
\begin{eqnarray}
E(k)=\;\half\!\!\!\!\!\!\!\!\sum_{k_- < |{\kk}|\leq k_+} \!\!\!\!\!\! |\hat{\BB}({\kk})|^2,\quad
H(k)=\;\Rey\!\!\!\!\!\!\!\!\sum_{k_- < |{\kk}|\leq k_+} \!\!\!\!\!\!
\hat{\AAA}\cdot\hat{\BB}^\ast,\;
\label{EHspec}
\end{eqnarray}
where $k_\pm=k\pm\delta k/2$ and $\delta k=2\pi/d$ is the wavenumber
increment and also the smallest wavenumber
\EQ
k_1\equiv\delta k=2\pi/d
\EN
in our cube of side length $d$.
The helicity spectrum $H(k)$ is not to be confused with
the electron scale height $H_e$.
We also compute the corresponding magnetic energy and helicity
transfer spectra \citep{Rem14}
\begin{eqnarray}
T_E(k)=-\Rey\!\!\!\!\!\!\!\!\!\sum_{k_- < |{\kk}|\leq k_+} \!\!\!\!\!\!\!\!
\widehat{\JJ}_{\kk}\cdot\widehat{(\JJ\times\BB)}_{\kk}^\ast,
\label{TEk}
\end{eqnarray}
\vspace{-3mm}
\begin{eqnarray}
T_H(k)=-\Rey\!\!\!\!\!\!\!\!\!\sum_{k_- < |{\kk}|\leq k_+} \!\!\!\!\!\!\!\!
\widehat{\BB}_{\kk}\cdot\widehat{(\JJ\times\BB)}_{\kk}^\ast.
\label{THk}
\end{eqnarray}
In the case when $\ff\neq\nullvector$, there are also source terms
$S_E(k)$ and $S_H(k)$ that are defined analogously to $T_E(k)$ and
$T_H(k)$, but with $\JJ\times\BB$ being replaced by $\ff$.
The magnetic energy and helicity spectra then obey
\begin{eqnarray}
\frac{\partial}{\partial t}E(k,t)=2T_E(k,t)-2\eta k^2 E(k,t)+2S_E(k,t),\;
\label{dEkdt}
\end{eqnarray}
\vspace{-3mm}
\begin{eqnarray}
\frac{\partial}{\partial t}H(k,t)=2T_H(k,t)-2\eta k^2 H(k,t)+2S_H(k,t).\;\,
\label{dHkdt}
\end{eqnarray}
The mean magnetic energy and helicity densities are defined as
${\cal E}=\bra{\BB^2}/2$ and ${\cal H}=\bra{\AAA\cdot\BB}$
in terms of the magnetic energy and helicity spectra as
\begin{equation}
\int E(k,t)\,dk={\cal E}(t),\quad
\int H(k,t)\,dk={\cal H}(t).
\end{equation}
The rms magnetic field can be obtained through $\Brms=(2{\cal E})^{1/2}$.
We define the magnetic correlation length $\xi$ as
\begin{equation}
\xi(t)=\left.\int k^{-1} E(k,t)\,dk\right/\!\!\int E(k,t)\,dk.
\label{xi_def}
\end{equation}
We define the instantaneous exponents describing
the growth of $\xi(t)$ and the decay of ${\cal E}(t)$ as
\begin{equation}
q(t)=d\ln\xi/d\ln t,\quad
p(t)= -d\ln{\cal E}/d\ln t.
\end{equation}
Those play important roles in describing the nature of the turbulence
in different cases \citep[cf.][]{BK17}.
To quantifying inverse cascading, we use a variable similar to $p$,
but now for the large-scale magnetic field only.
Because we expect the magnetic field to increase at large scales,
it will be defined with a plus sign, i.e.,
\begin{equation}
p_{\rm LS}(t)=d\ln{\cal E}_{\rm LS}/d\ln t,
\end{equation}
where ${\cal E}_{\rm LS}(t)=\int_{k_1}^{3k_1}E(k,t)\,dk$, which is an
arbitrarily chosen compromise between relying only on a single wavenumber
(just $k_1$) and some other weighted average that takes the entire
spectrum into account, but emphasizes the low wavenumbers.

\subsection{Invariance under rescaling}

In connection with decaying hydrodynamic and MHD turbulence studies,
\cite{Ole97} was the first to make use of the invariance
of the MHD equations under rescaling of space and time coordinates,
along with a corresponding rescaling of the other dependent variables.
A similar procedure applies analogously to \Eq{dAdt}, which is invariant
under the following rescaling:
\begin{eqnarray}
\label{ReScaling}
&&t=\tau t',\quad
\xx=\tau^q \xx',\quad
\eta=\tau^{2q-1} \eta',\\
&&\AAA=\tau^{3q-1} \AAA',\quad
\BB=\tau^{2q-1} \BB',\quad
\JJ=\tau^{q-1} \JJ'.\nonumber
\end{eqnarray}
Inserting these variables into \Eq{dAdt}, the resulting equation in
the primed quantities has the same form as \Eq{dAdt} in its original
formulation.
This requires that $\eta\propto t^r$ where $r=2q-1$.
For $q<1/2$, $r$ is negative so $t^r$ becomes singular for $t\to0$.
Therefore, we use in such cases
\begin{eqnarray}
\eta(t)=\eta_0 [\max(1,t/t_0)]^r,
\label{etat_formula}
\end{eqnarray}
where $t_0$ is the time below which $\eta$ is assumed fixed.

It should be noted that the scaling in \Eq{ReScaling} is different
from that found in MHD, where, ignoring the density factor,
$[\BB]=[\xx]/[t]\propto\tau^{q-1}$.
This difference is significant and results in new relationships between
$p$ and $q$.
We return to this in the next section.

In view of the limited dynamical range available in numerical simulations
of decaying turbulence, the use of a time-dependent $\eta$ implies
significant computational advantages in that a suitably defined Lundquist
number is then approximately constant over much of the duration of
the run.
It allows us to identify selfsimilar scaling properties.
In particular, since both $\Brms$ and $\eta$ decay in time, we can
characterize a ``typical'' $\Brms$ by specifying the temporal average
of their normalized ratio, namely
\EQ
\tilde{B}_{\rm rms}\equiv\bra{\Brms/(en_e\mu_0\eta)}.
\EN
Similarly, the magnetic dissipation can be expressed in a similar
fashion as
\EQ
\tilde{\epsilon}\equiv\bra{\epsilon/(e^2 n_e^2 \mu_0 \eta^3/\xi^2)},
\EN
where not only $\eta$ decreases, but $\xi$ increases such that the
normalization factor of $\epsilon$ decreases in a similar fashion.
Furthermore, since $\xi$ increases with time, the averaged ratio
\EQ
\tilde{\eta}\equiv\bra{t\eta/\xi^2}
\EN
is another quantity that we quote for our runs to characterize the
effective value of $\eta$.

The definitions of $\tilde{B}_{\rm rms}$, $\tilde{\epsilon}$, and
$\tilde{\eta}$ remain somewhat problematic in that they are averages
over ratios that can still show a residual trend, so the result
depends on the time interval of averaging.
However, for the purpose of this paper, we only want to provide a first
orientation.
We should keep this caveat in mind when those values are quoted below.

\subsection{Dimensional argument for inertial range scaling}
\label{DimensionalArgument}

\cite{Bis+96} where the first to suggest a $k^{-7/3}$ inertial range
spectrum scaling based on an energy transfer rate proportional to the
cube of the electron velocity potential.
\cite{WH09} proposed a slightly different scaling proportional to
$k^{-5/2}$, but did not suggest any phenomenology for that.
It is clear that dimensional arguments cannot emerge when one expresses
the magnetic field in velocity units, as is usually done in MHD
\citep[see also][]{CL09}.
A physically more meaningful normalization for the Hall cascade is in
terms of diffusivity units by noting that $\BB/en_e\mu_0$ (where $\BB$
is here in Tesla) has dimensions of $\m^2\s^{-1}$.
In those units, $E(k)$ has dimensions $\m^5\s^{-2}$, and $\epsilon$
has dimensions $\m^4\s^{-3}$.
Making the ansatz
\EQ
E(k)=C_{\rm Hall}\epsilon^a k^b
\label{HallEq}
\EN
with exponents $a$ and $b$ and the dimensionless coefficient
$C_{\rm Hall}$, we find, on dimensional grounds, $a=2/3$ and $b=-7/3$,
which is consistent with the result of \cite{Bis+96}, although the
coefficient $C_{\rm Hall}$ has not previously been introduced in this
form, nor have estimates for its value been provided.
Results for $C_{\rm Hall}$ will be given below, separately for helical
and nonhelical turbulence.

In the time-dependent case, given that $E(k,t)$ has now different
dimensions than in MHD, we have to adapt the corresponding arguments of
\cite{Ole97} and \cite{BK17} for selfsimilar solutions.
If solutions are selfsimilar, the simultaneous dependence on $k$ and
$t$ can be captured by a function $\phi$, which depends only on the
scaled wavenumber $k\xi(t)$, such that the peak of the spectrum is
always at $k\xi(t)\approx1$.
In addition, the decrease of the amplitude of the spectrum with time is
compensated by the prefactor $\xi^{-\beta}$ with some exponent $\beta$,
so we have
\EQ
E\big(k\xi(t),t\big)=\xi^{-\beta}\phi(k\xi).
\label{EXIbeta}
\EN
We must require $E$ itself to be invariant under rescaling,
\EQ
E\to E'\tau^{5q-2}\propto\xi^{-\beta}\tau^{-\beta q}\phi(k\xi),
\EN
so we must require that $5q-2=-\beta q$, and therefore $\beta=2/q-5$.
This relation is similar to that in MHD, except that the 5 is then
replaced by 3.

Let us relate this now to the decay law for energy, which is of the
form ${\cal E}\propto t^{-p}$.
Since
\EQ
{\cal E}(t)=\int E(k,t)\,dk\propto \xi^{-\beta},
\label{EExiBeta}
\EN
and using $\xi\propto t^q$, we have $p=(1+\beta)q$, just like in MHD;
see Equation~(6) of \cite{BK17}.
This implies that, in the $pq$ diagram, selfsimilar solutions must lie
on the line
\EQ
p=2(1-2q)\quad\mbox{(selfsimilarity line)},
\EN
which is steeper than the corresponding line $p=2(1-q)$ in MHD.

As in MHD, the relevant values of $\beta$ and $q$ depend on the
physics governing the decay.
If the decay is governed by magnetic helicity \citep{Cho11}, which has now
dimensions $[x]^5/[t]^2$, then $q=2/5$ and $\beta=0$.
Whether or not there can be other relevant quantities in the nonhelical
case analogous to the Loitsiansky integral ($\beta=4$) or the Saffman
integral ($\beta=2$) is unclear.
In MHD, the case $\beta=1$ has been associated with the possibility
that some locally projected squared vector potential, $\AAA_\perp^2$,
may be conserved \citep{BKT15}.
Some of our simulations point to a possible relevance of lines with
$\beta=2$--$3$ in the $pq$ diagram.
In \Tab{pq_tab} we summarize the other associated coefficients for
several values of $\beta$.

The result ${\cal E}\propto \xi^{-\beta}$ in \Eq{EExiBeta} is a
consequence of integrating $\phi(k\xi)$ over all $k\xi$, which gives
just a number, leaving only $\xi^{-\beta}$ outside the integral.
In the calculation of ${\cal E}_{\rm LS}$, however, only a fixed $k$
range matters, so the result depends on the slope of $\alpha$ of the
subinertial range and the slope $\beta$ of the envelope, and therefore
we expect $p_{\rm LS}=(\alpha-\beta)q$.
Using $\beta=2/q-5$ and $\alpha=5$ (see below), this gives
$p_{\rm LS}=10q-2$.
For the helical case with $q=2/5$, we therefore expect $p_{\rm LS}=2$.

\subsection{Numerical simulations}

For our numerical simulations we use the {\sc Pencil Code}
(\url{https://github.com/pencil-code})
which is a public MHD code that is particularly
well suited for simulating turbulence.
In practice, both $B_0$ and $\eta_0$ were varied, but the decisive
control parameter is just a ratio $B_0/en_e\mu_0\eta$.
However, in all the tables and plots presented below, we express the
results in normalized form.
The numerical resolution is $1024^3$ meshpoints in most of the cases
presented below.
A summary of all simulations is given in \Tabs{Tmodels}{TmodelsIII}.
Here, $t_2$ denotes the end time of the simulation.
In Runs~A--F, the start time is $t=0$, but our data analysis commences
at $t=t_1$.
The different values of $t_2$ are partly explained by the different
speeds at which $\xi$ grows and also the different numerical time steps,
making runs with stronger magnetic field less economic to run.
We usually spend 1--7~days per run on 1024 processors on a Cray~XC40,
where a time step takes about $0.4\s$.
The run directories for simulations are publicly available; see
\cite{HallCascade}.

\section{Results}

\subsection{Inertial range for stationary case}

\begin{figure}\begin{center}
\includegraphics[width=\columnwidth]{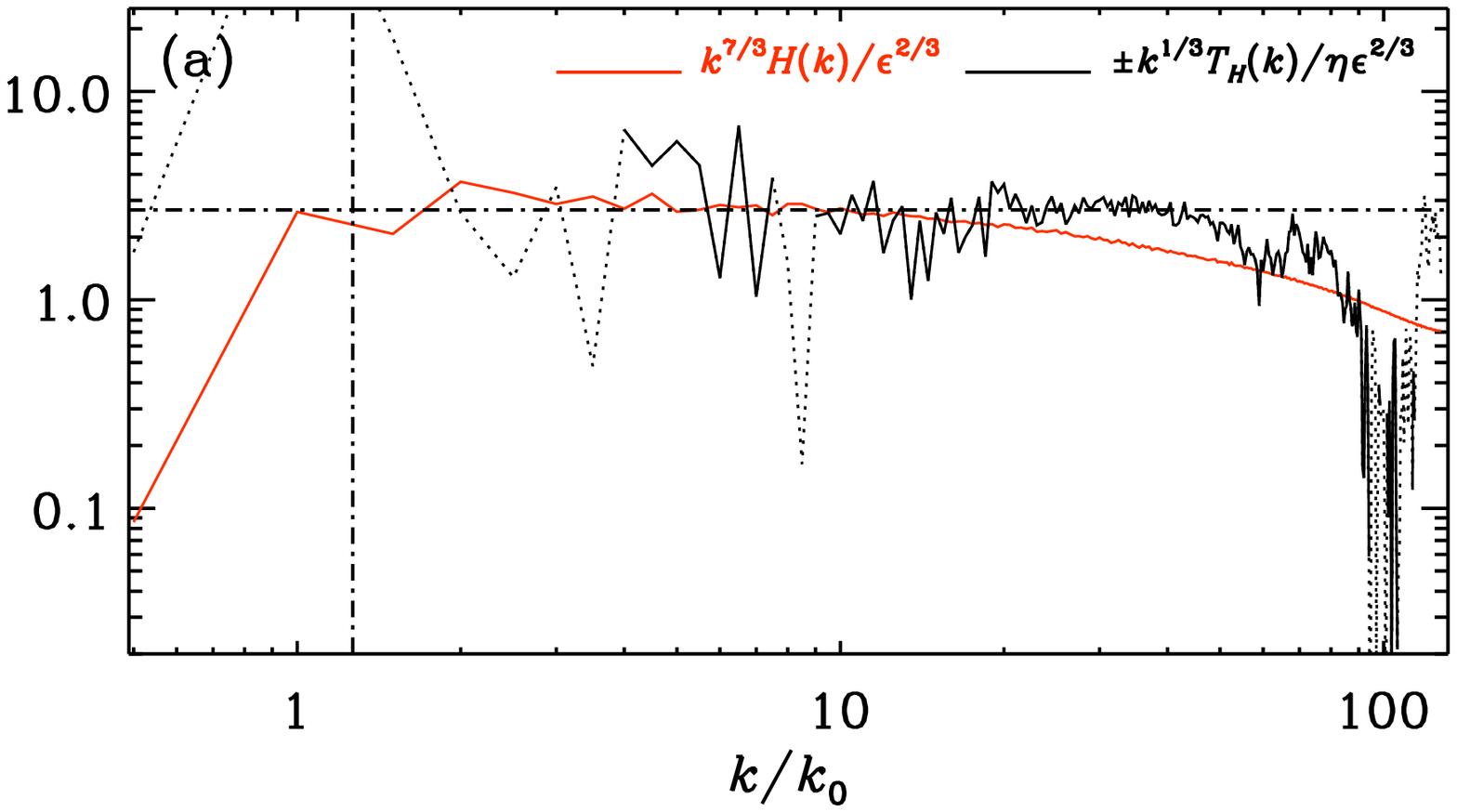}
\includegraphics[width=\columnwidth]{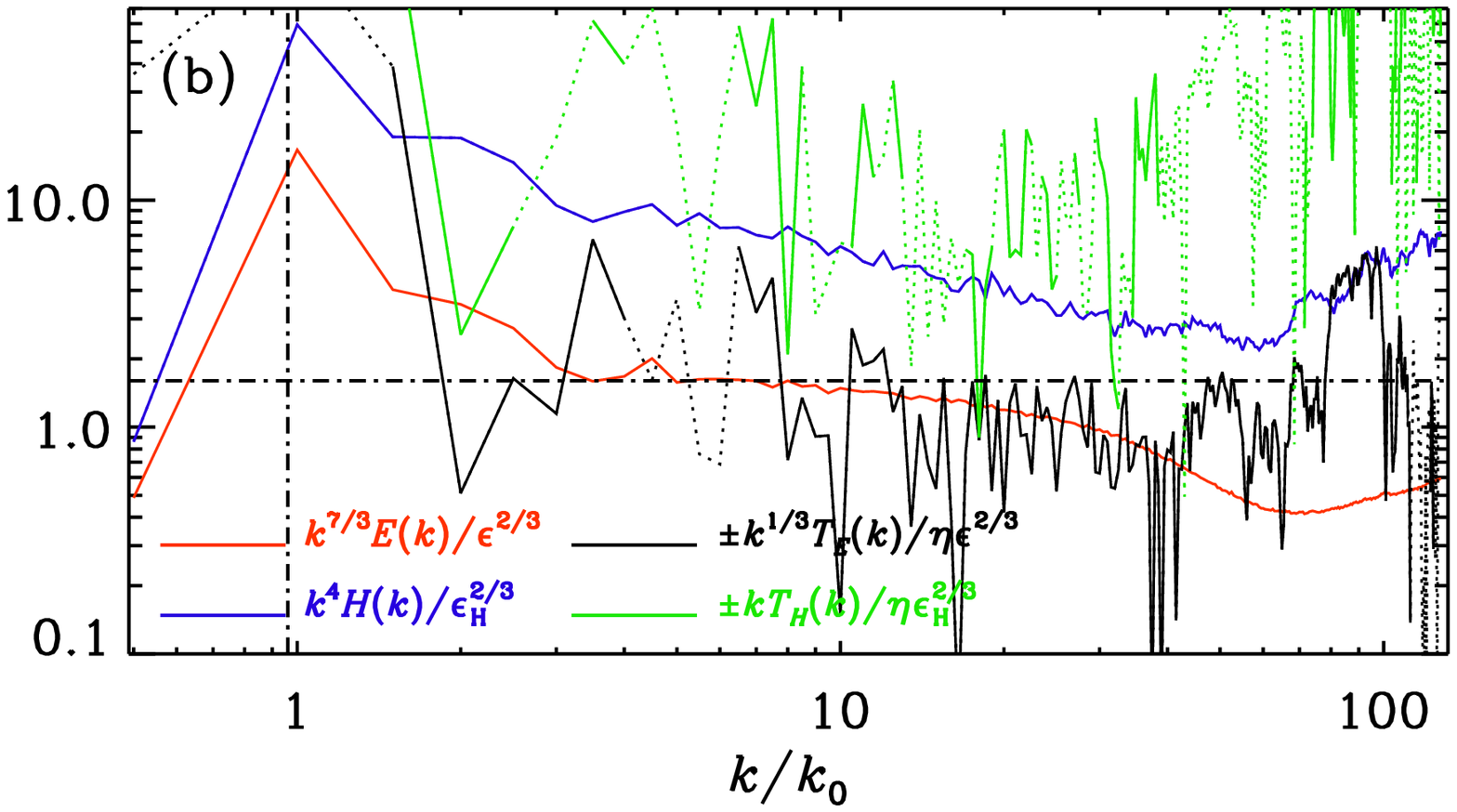}
\end{center}\caption[]{
Stationary Hall cascade for Run~II without helicity (a)
and Run~III with helicity (b).
Solid (dotted) parts of the lines denote positive (negative) values.
}\label{pkt_one_F512a_cont}\end{figure}

We begin by making contact with previous work and verify that the
expected $k^{-7/3}$ inertial range scaling is obtained in
the stationary case.
To simulate this, we invoke the forcing term $\ff$ during the
entire time of the simulation.
We choose $k_0=2k_1$.
The resulting spectra, compensated by $\epsilon^{-2/3}k^{7/3}$, are
shown in \Fig{pkt_one_F512a_cont} for nonhelical and helical forcings.
Note that the $k^{-7/3}$ scaling is reproduced in both cases.
Furthermore, there is no (or not the usual) diffusive subrange.
The lack of a diffusive subrange was already emphasized by \cite{WH10},
which they ascribed to nonlocal spectral energy transfer.
This lack of a diffusive subrange is related to the fact that the highest
derivative in the linear and nonlinear terms in \Eqs{dAdt}{dAdt2} is
the same, i.e., both terms are proportional to $\JJ$.
Earlier simulations \citep{Bis+96,Bis+99,CL04} used hyperviscosity,
which leads to an artificial dissipative cutoff, precluding any statements
about a naturally occurring cutoff.

\begin{table}\caption{
Selfsimilarity parameters for different values of $\beta$.
\vspace{-3mm}
}\vspace{12pt}\begin{tabular}{ccrcl}
$\beta$ & $q$ & $p\quad$ & $r$ & comment \\
\hline
0 & $2/5=0.40$ & $2/5=0.40$ &  $-0.20$ & $\bra{\AAA\cdot\BB}=\const$\\
1 & $2/6=0.33$ & $4/6=0.86$ &  $-0.33$ & $\bra{\AAA_\perp^2}=\const$ ?\\
2 & $2/7=0.29$ & $6/7=0.86$ &  $-0.43$ & Saffman-type scaling\\
3 & $2/8=0.25$ & $8/8=1.00$ &  $-0.50$ & \\
4 & $2/9=0.22$ &$10/9=1.11$ &  $-0.56$ & Loitsiansky-type scaling\\
\label{pq_tab}\end{tabular}\end{table}

\begin{table*}\caption{
Parameters for the unstratified models.
}\vspace{12pt}\begin{tabular}{ccccccrlrlrccccl}
Run & $f_0$ & $B_0/[B]$ & $\Brms^{(0)}/[B]$ & $\sigma_0$ & $-r$ &
$k_0/k_1\!\!\!$ & $t_1/[t]$ & $t_2/[t]$ & $\tilde{\eta}\;\;$ &
$\tilde{B}_{\rm rms}$ & $\;\;\tilde{\epsilon}$ & $p_{\rm LS}$ & $p$ & $q$ & comment\\
\hline
A&0& 200 & 600  &  0   &0.43&180&0.2   &4600&0.13  &   80&$2\times10^4$& 0.5 & 0.9 & 0.3 & $\beta=3\to2$ \\
B&0&2000 &6000  &  0   &0.43&180&0.1   & 500&0.024 &  600&$3\times10^6$& 1.4 & 0.8 & 0.3 & $\beta\approx2$\\
C&0&2000 &6000  &$10^{-3}$&0.43&180&0.1  &1000&0.020 &  800&$6\times10^6$& 1.5 & 0.4 & 0.4 & $\beta=2\to0$ \\
C'&0&3000&10000 &$10^{-3}$&0.20&180&0.1   & 100&0.040 &  300&$5\times10^5$& 1.8 & 0.5 & 0.4 & $\beta=2\to0$ \\
D&0&2000 &6000  &$10^{-2}$&0.43&180&0.1  &  50&0.012 & 1500&$3\times10^7$& 2.5 & 0.4 & 0.4 & $\beta=2\to0$ \\
E&0&2000 &8000  &   1   &0.43&180&0.001&  10&0.003 & 6000&$2\times10^8$& 2.5 & 0.4 & 0.4 & $\beta=0$ \\
F&0& 200 &1000  &   1   & 0 &180&0.01  &  16&0.02  & 1000&$6\times10^6$& 1.3 & 0.6 & 0.3 & $\beta=0$ \\
a&$4\times10^{-4}$&150& 350 &0&0.43&60&0.02  &1800&0.09  &  130&$7\times10^4$& 0.8 & 0.9 & 0.3 & $\beta\approx2$ \\
b&$4\times10^{-3}$&1500&3500 &0&0.43&60&0.02  &  75&0.015 & 1100&$1\times10^7$& 1.3 & 1.0 & 0.3 & $\beta\approx2$ \\
c&$3\times10^{-2}$&1500&4000 &1&0.43&60&0.02  &   4&0.004 & 5000&$2\times10^8$& 2.2 & 0.4 & 0.4 & $\beta=0$\\
\label{Tmodels}\end{tabular}
\tablenotemark{
The resolution is $1024^3$ mesh points for Runs~A--F and $512^3$ mesh
points for Runs~a--c.
}\end{table*}

\begin{table}\caption{
Model parameters for stationary turbulence.
}\vspace{12pt}\begin{tabular}{cccccccc}
Run & $f_0$ & $\sigma_0$ & $\tilde{B}_{\rm rms}$ & $\tilde{\epsilon}_{\rm rms}$ & $C_{\rm Hall}$ \\
\hline
  I & $4\times10^{-2}$ & 0 &1450 & $15\times10^7$ & 4.5 \\
 II & $4\times10^{-3}$ & 0 & 300 &  $2\times10^6$ & 2.7 \\
III & $4\times10^{-3}$ & 1 & 360 &  $7\times10^5$ & 1.6 \\
\label{TmodelsIII}\end{tabular}
\tablenotemark{
The resolution is $512^3$ mesh points and $k_0/k_1=2$ in all three cases.
}\end{table}

By compensating the spectra with $\epsilon^{-2/3}k^{7/3}$,
the value of $C_{\rm Hall}$ in \Eq{HallEq} can be read off from
\Fig{pkt_one_F512a_cont} as the height of the plateau.
We find $C_{\rm Hall}\approx2.7$ in the nonhelical case (Run~II)
and $C_{\rm Hall}\approx1.6$ in the helical case (Run~III);
see \Tab{TmodelsIII}.
For Run~I with stronger forcing, however, we find $C_{\rm Hall}\approx2.7$,
but this could be because the resulting magnetic field strength is here
too large for the numerical resolution, so the value of $\epsilon$ could
be underestimated and therefore the compensated value appears too large.
Note also that $T_E(k)$ has been scaled by
$\eta_0^{-1}\epsilon^{-2/3}k^{1/3}$, which allows us to see
that \Eq{dEkdt} is approximately obeyed.
In the helical cases, the current helicity displays a forward cascade
\citep{BS05b}.
We therefore write
$k^2H(k,t)=C_{\rm Hall}^{\rm hel}\epsilon_{\rm H}^{2/3}k^{-2}$,
which follows from dimensional arguments analogous to those for $E(k,t)$.
Here, $C_{\rm Hall}^{\rm hel}$ is a new coefficient and
$\epsilon_{\rm H}=2\eta\int k^2H(k,t)\,dk$ is the magnetic helicity
dissipation.
However, our simulation results for $k^4H(k,t)$ do not show a plateau,
so we refrain here from pursuing this question further.

\begin{figure}\begin{center}
\includegraphics[width=\columnwidth]{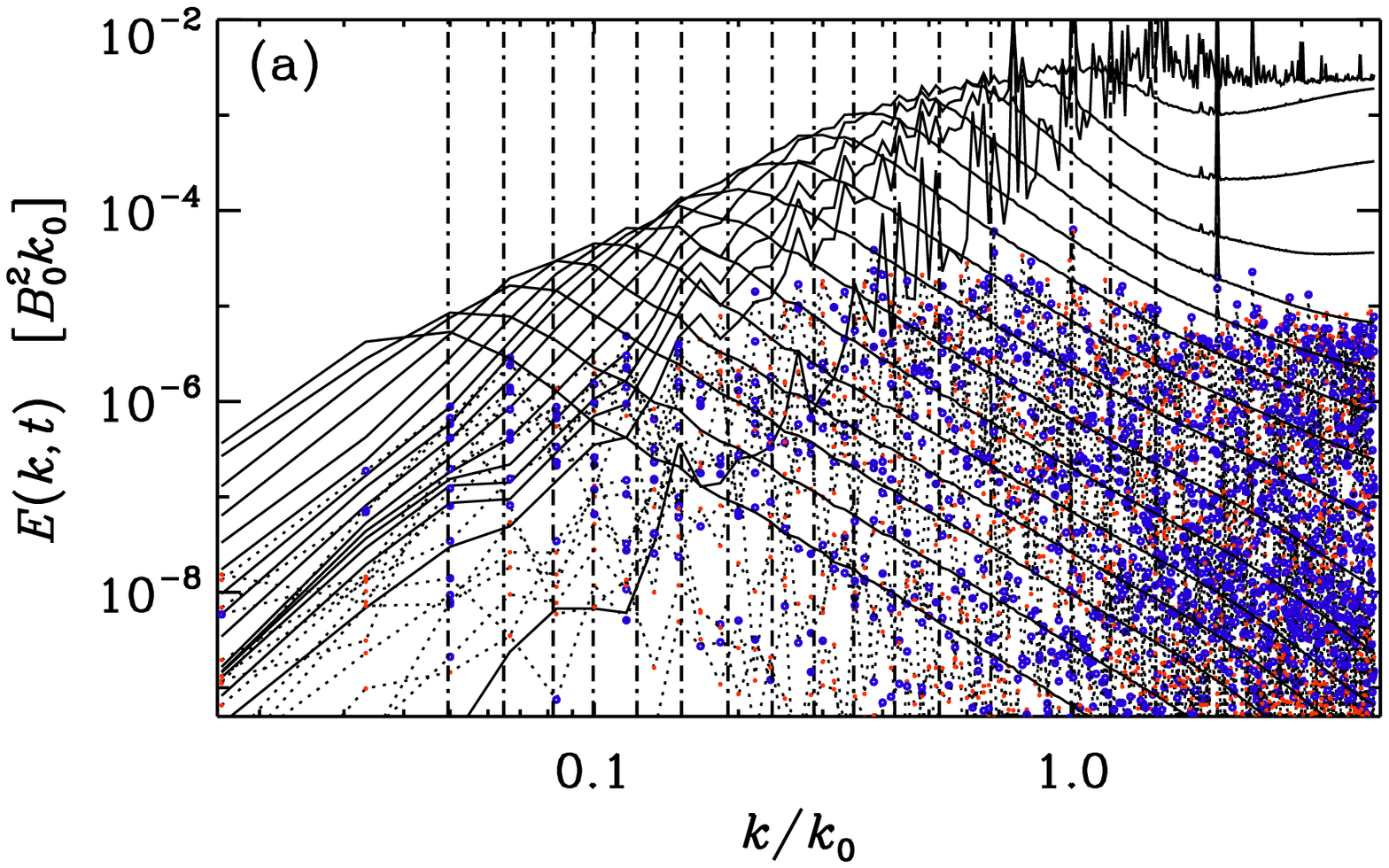}
\includegraphics[width=\columnwidth]{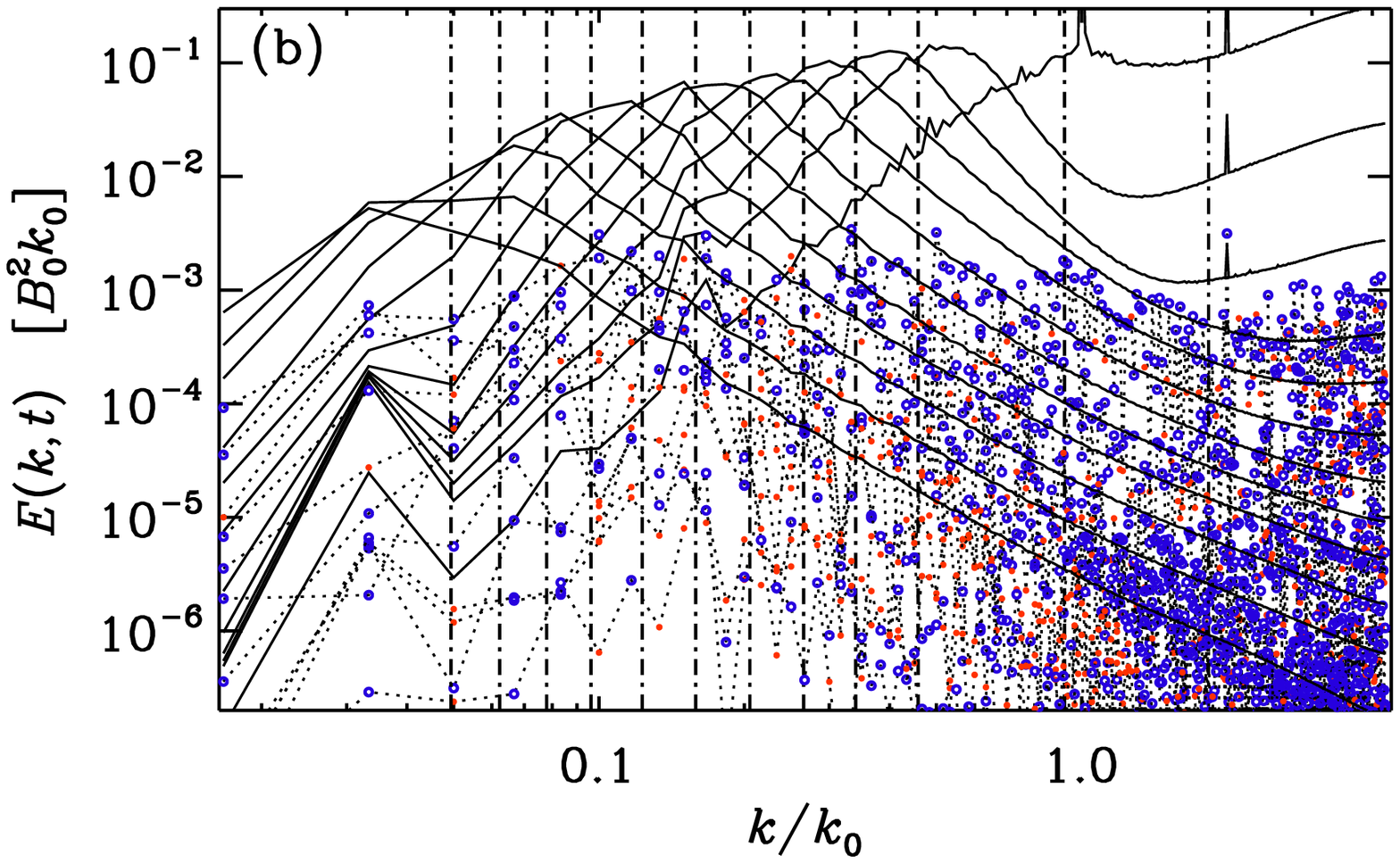}
\end{center}\caption[]{
$E(k,t)$ (solid black lines) and 
$H(k,t)$ (dotted lines) with positive (negative) values indicated
by red (blue) closed (open) symbols for the nonhelical Runs~a and b.
The vertical dash-dotted lines show the positions where $k\xi=1$.
}\label{pkt_few_f512c}\end{figure}

\begin{figure}\begin{center}
\includegraphics[width=\columnwidth]{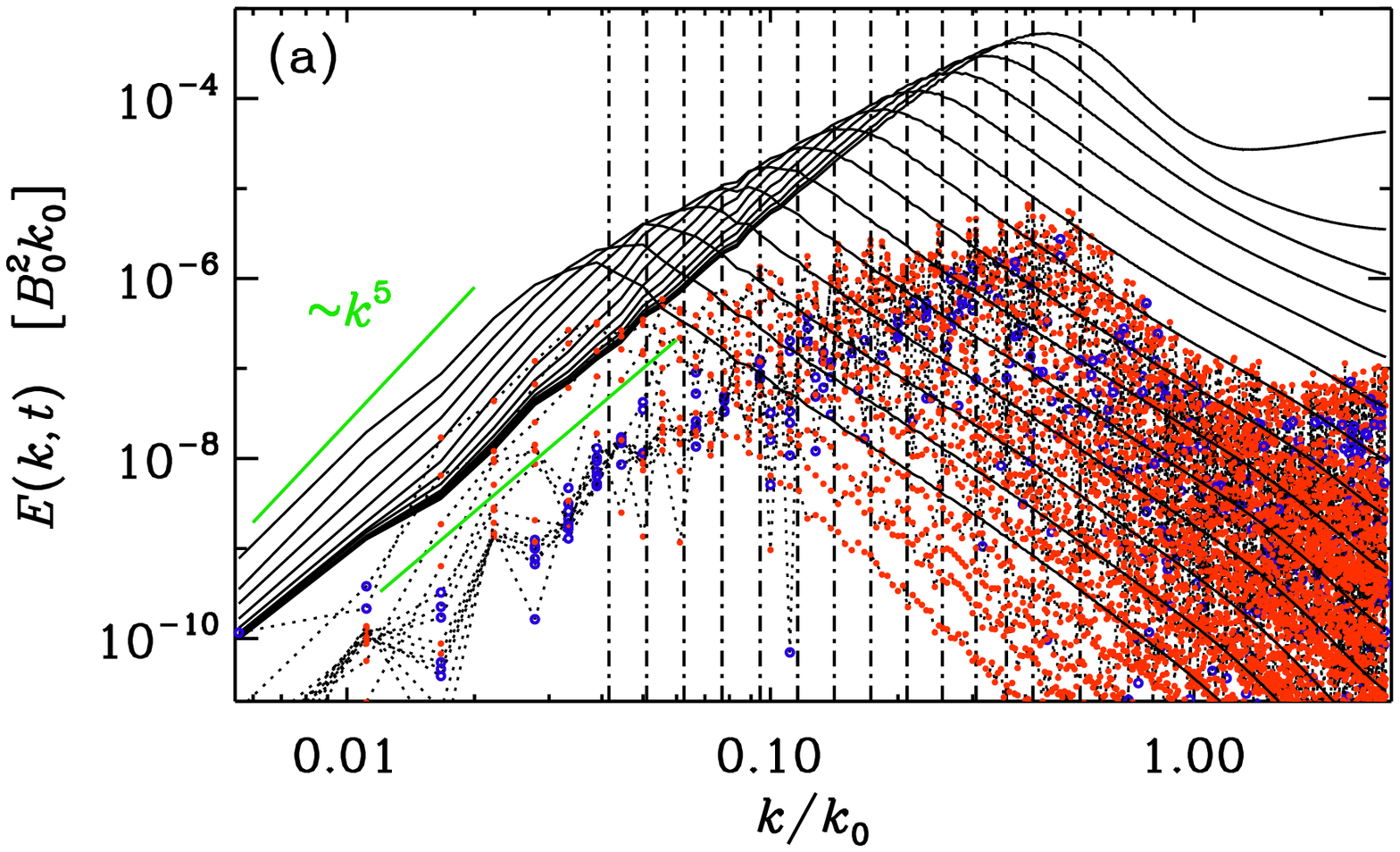}
\includegraphics[width=\columnwidth]{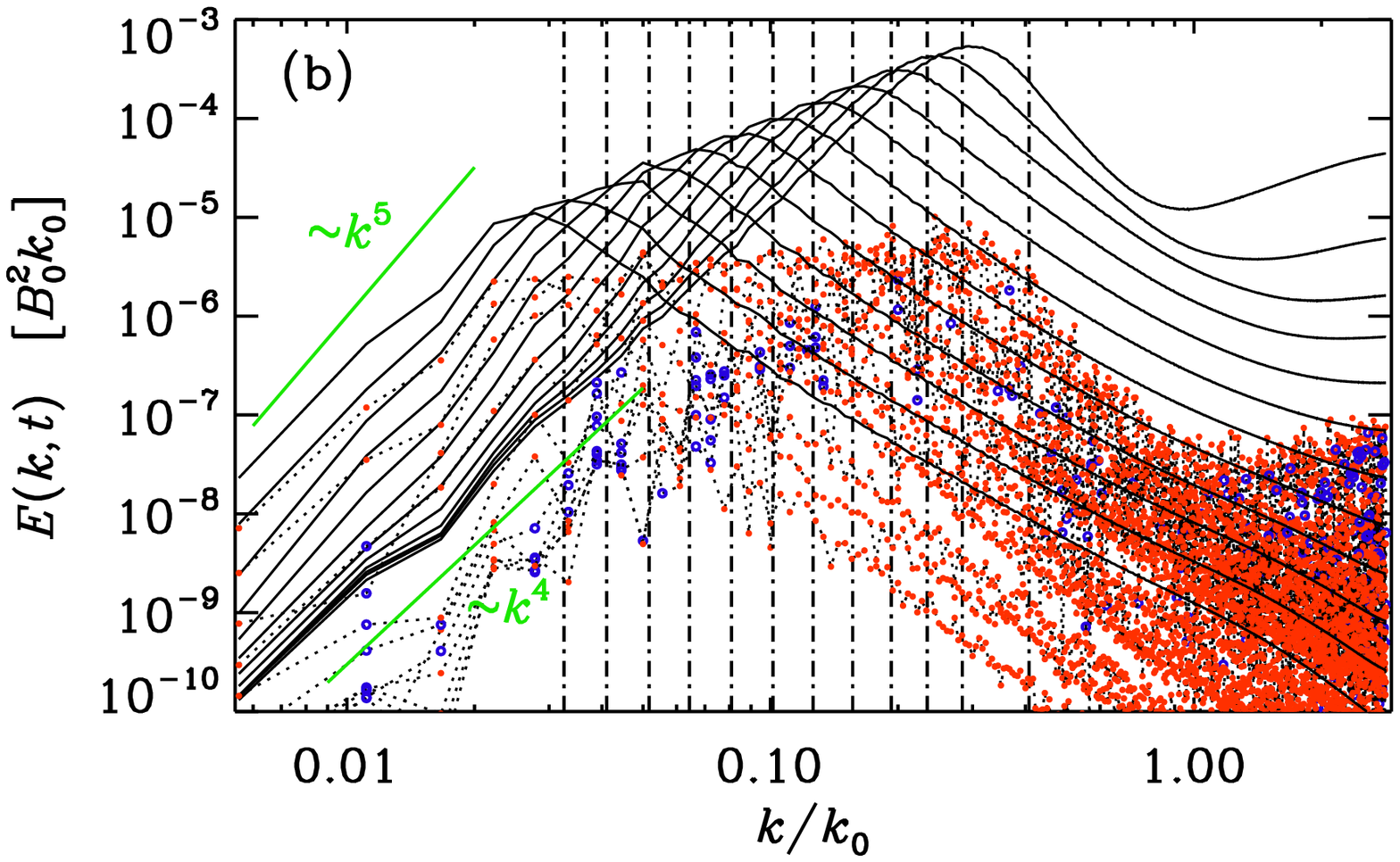}
\end{center}\caption[]{
Similarly to \Fig{pkt_few_f512c}, but for (a) Run~A and (b) Run~B
with weak and strong fields.
}\label{pkt_few_Hf0_t2em5_k180a}\end{figure}

\subsection{Nonhelical decay}

We now consider a nonhelical initial magnetic field.
The case where an initial magnetic field is obtained via short-term
monochromatic driving is shown in \Fig{pkt_few_f512c}, where we
present the resulting spectra at different times.
To see whether the magnetic helicity plays a role in our simulations,
we plot $|kH(k,t)/2|$ together with $E(k,t)$.
This representation is useful because of the realizability condition,
which states that
\EQ
|kH(k,t)/2|\leq E(k,t),
\label{RealSpec}
\EN
so we see immediately at which wavenumbers the inequality is closest
to saturation.
Since $H(k,t)$ can have either sign, we use red (blue) symbols to indicate
positive (negative) values.
We see that, at early times, $|kH(k,t)/2|$ is much smaller than $E(k,t)$.
This changes at later times after $E(k,t)$ has dropped by several orders
of magnitude.
It is possible that the realizability condition limits further decay of
$E(k,t)$, even though $H(k,t)$ does not have a definite sign.

\begin{figure}\begin{center}
\includegraphics[width=\columnwidth]{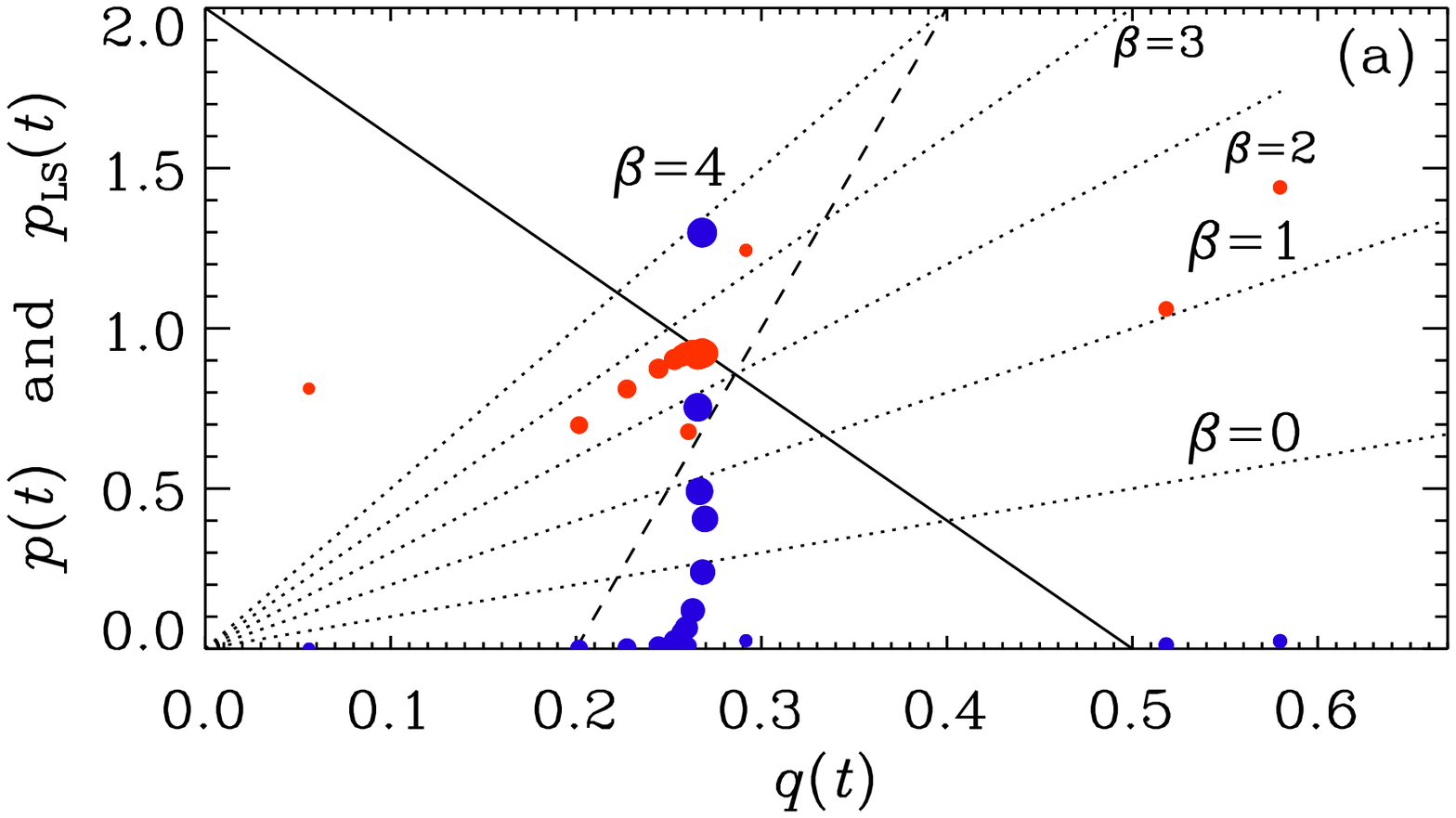}
\includegraphics[width=\columnwidth]{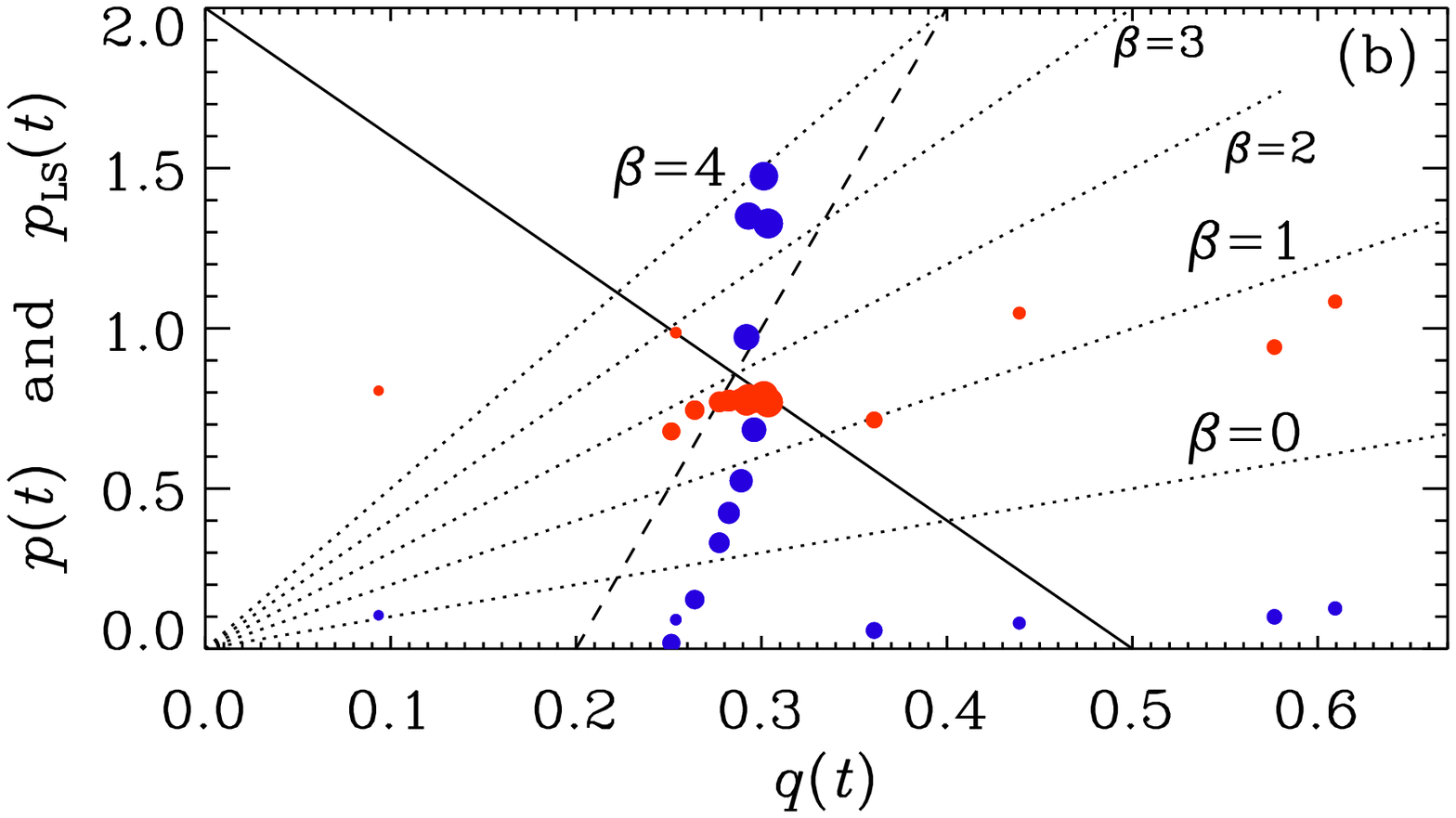}
\end{center}\caption[]{
$pq$ diagram showing $p$ (red symbols) and $p_{\rm LS}$ (blue symbols)
versus $q$ for Runs~A and B in panels (a) and (b), respectively.
Larger symbols indicate later times.
The selfsimilarity line (solid) and the $p_{\rm LS}=10q-2$ line
(dashed) of \Sec{DimensionalArgument} are also shown.
}\label{pq_Hf0_t2em5_k180c}\end{figure}

By comparison, the case with an initial power law spectrum is shown in
\Fig{pkt_few_Hf0_t2em5_k180a} for weak and strong magnetic fields.
It turns out that, depending on the strength of the initial magnetic
field, there is always a certain amount of inverse transfer, i.e.,
the spectral energy increases with time at small $k$, so $p_{\rm LS}>0$.
This confirms earlier findings by \cite{CL09}, \cite{WH09}, and \cite{Cho11}.

In \Fig{pq_Hf0_t2em5_k180c} we show the $pq$ diagram for Run~A.
Note the convergence of the point $(p,q)$ toward the selfsimilarity line
with $p\approx0.9$ and $q\approx0.3$.
We have chosen to plot $p_{\rm LS}$ in the same plot, although it
reflects different physics not related to $\beta$.
It simply allows us to obtain a visual impression of how $p_{\rm LS}$
changes.
In the present case, we see that $p_{\rm LS}$ approaches the value 0.5.

\subsection{Approach to $k^5$ scaling}

Our initial conditions usually have a $k^4$ subinertial range spectrum.
In the cosmological context, such a spectrum is motivated by causality
requirements for early times \citep{DC03}.
It follows from a $\delta$-correlated magnetic vector potential,
so the shell-integrated spectrum of $\AAA$ corresponds to that of white
noise and that of $\BB$ corresponds to that of blue noise.

Looking at \Fig{pkt_few_Hf0_t2em5_k180a}(b), we see that the $k^4$
spectrum gradually evolves toward $k^5$.
This steepening is rather remarkable and has never been seen in MHD.
In MHD, by contrast, it is known that, in the presence of magnetic
helicity, a shallower initial spectrum involves gradually toward a $k^4$
spectrum; see Figure~3(a) of \cite{BK17}.

\begin{figure}\begin{center}
\includegraphics[width=\columnwidth]{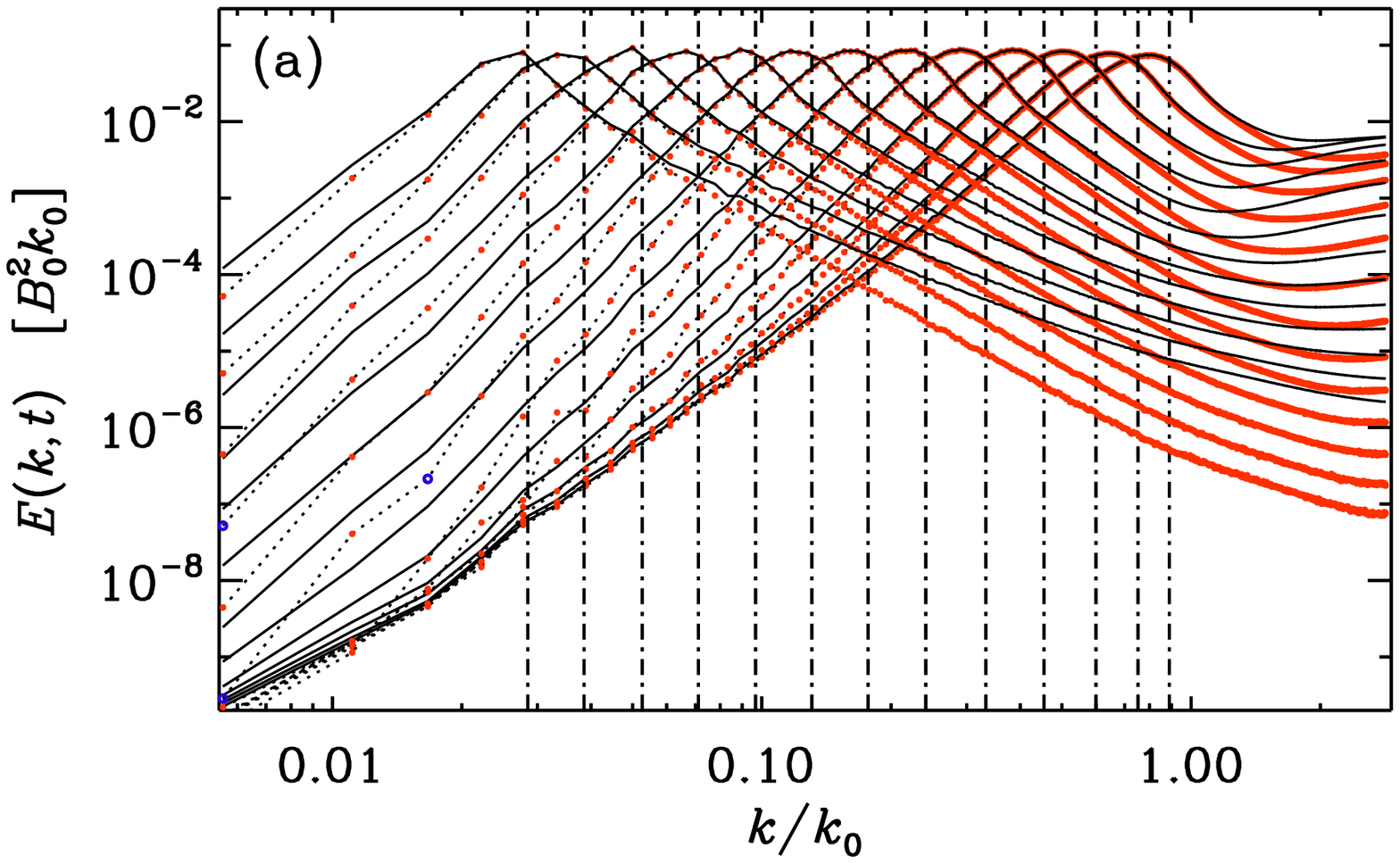}
\includegraphics[width=\columnwidth]{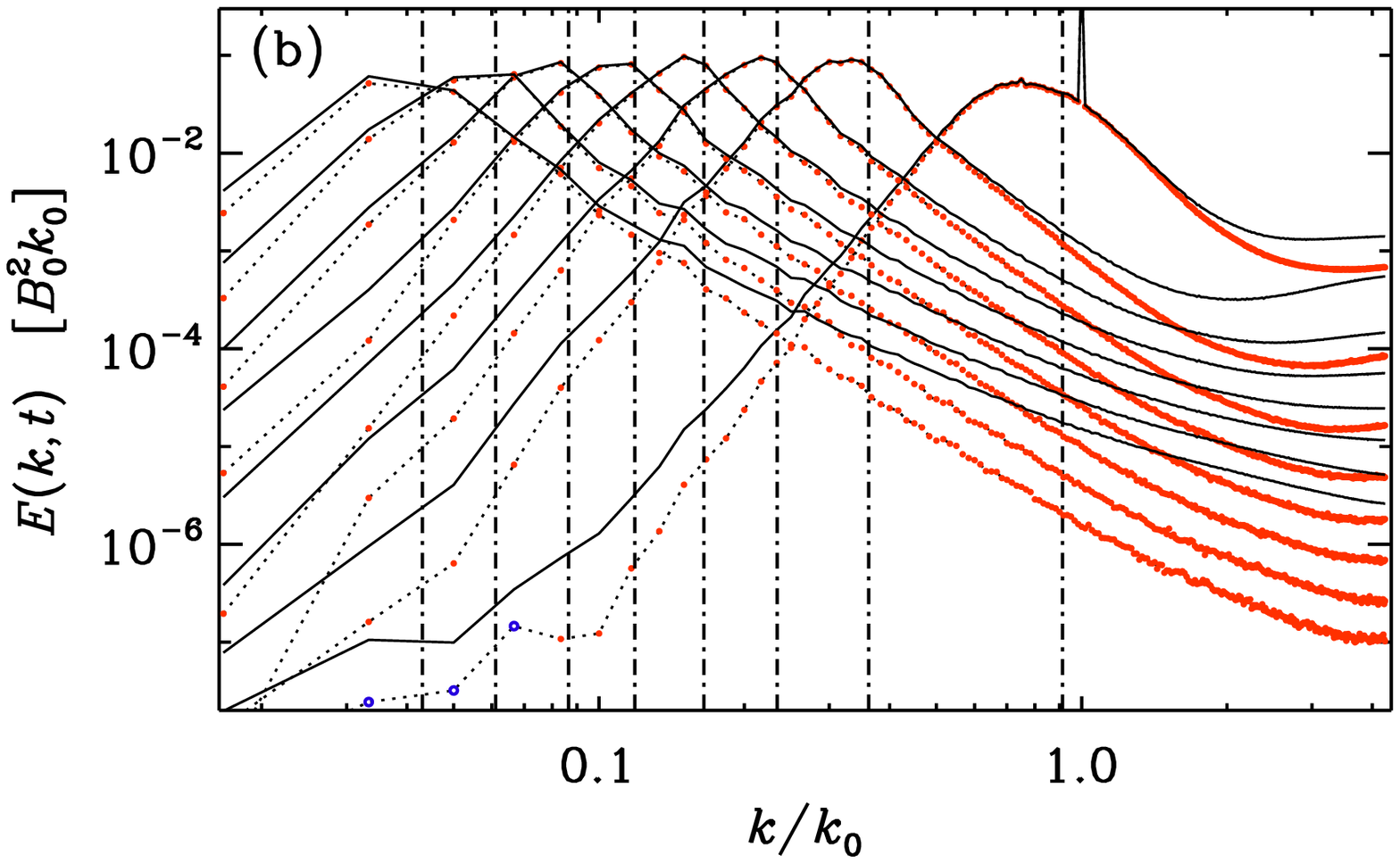}
\end{center}\caption[]{
Similarly to \Fig{pkt_few_f512c}, but for (a) Run~E and (b) Run~c,
which have different initial conditions, but are both fully helical.
}\label{pkt_few_Hf1_t2em5_k180a}\end{figure}

Inverse transfer, on the other hand, has been seen in nonhelical MHD
with strong magnetic fields \citep{BKT15}, but here the effect is much
more pronounced.
This can be qualitatively explained by the nonlinearity of \Eq{dAdt},
because we have seen that the nonhelical inverse transfer is stronger
for a stronger magnetic fields.
Therefore, the peak of the magnetic energy spectrum, where the field is
stronger, is expected to move faster toward lower wavenumbers than the
lower parts where the field is weaker.
This is seen in \Fig{pkt_few_Hf0_t2em5_k180a}(a), where the magnetic
field is weaker and the spectrum remains somewhat shallower than $k^5$.

\subsection{Fully helical initial fields}

Next, we demonstrate the effect of finite magnetic helicity.
It is well known that its presence constrains the mean magnetic energy
density from below, so that
\EQ
{\cal E}(t)\ge|{\cal H}|/2\xi(t),
\label{RealSpac}
\EN
which is similar to the spectral realizability condition mentioned above.
Since ${\cal H}$ is nearly constant, ${\cal E}(t)$ can only decrease if
$\xi(t)$ increases at the same rate, and therefore $p=q$.

\begin{figure}\begin{center}
\includegraphics[width=\columnwidth]{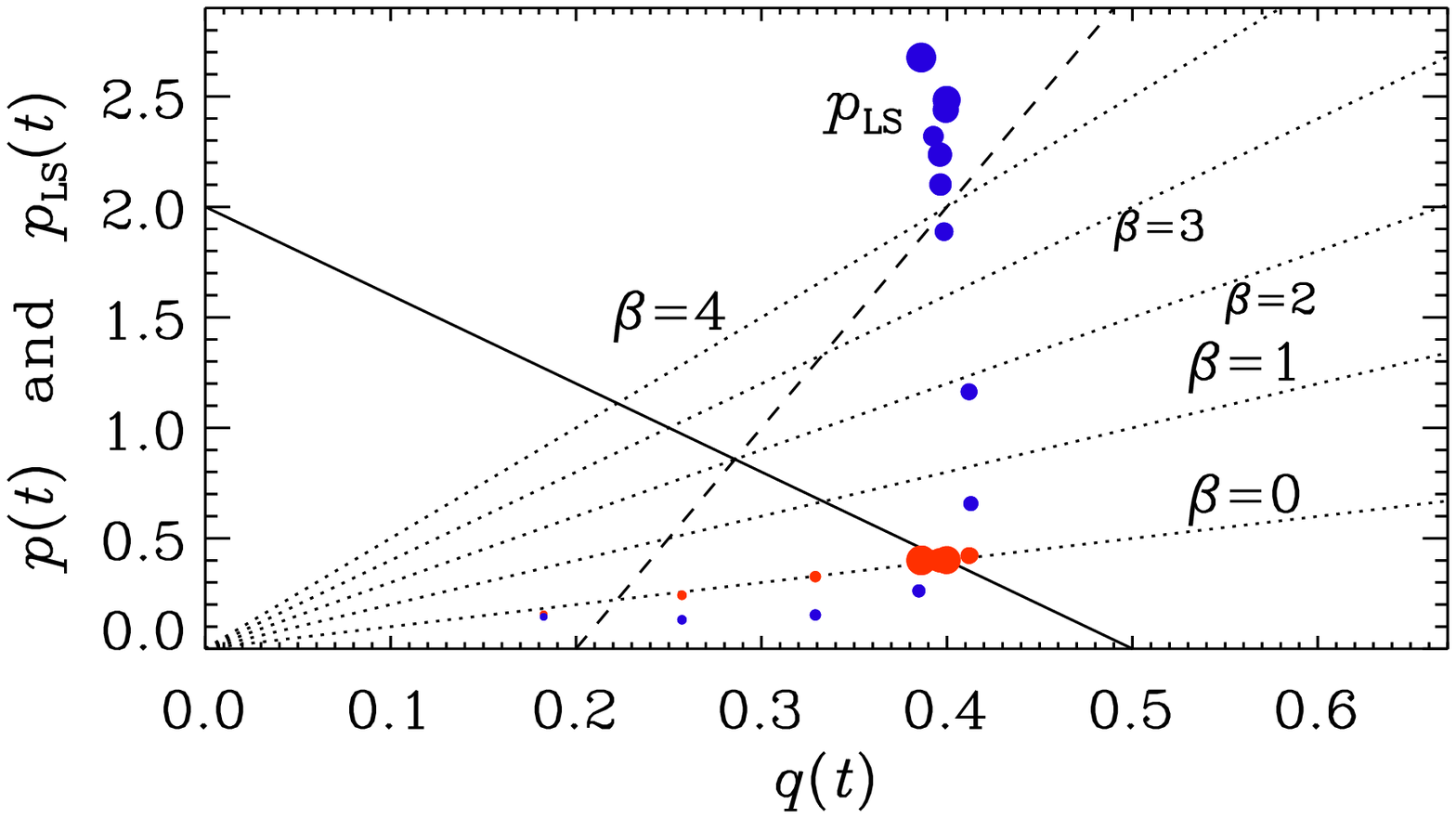}
\end{center}\caption[]{
Similar to \Fig{pq_Hf0_t2em5_k180c}, but for Run~E.
}\label{pq_Hf1_t2em5_k180a}\end{figure}

\begin{figure}\begin{center}
\includegraphics[width=\columnwidth]{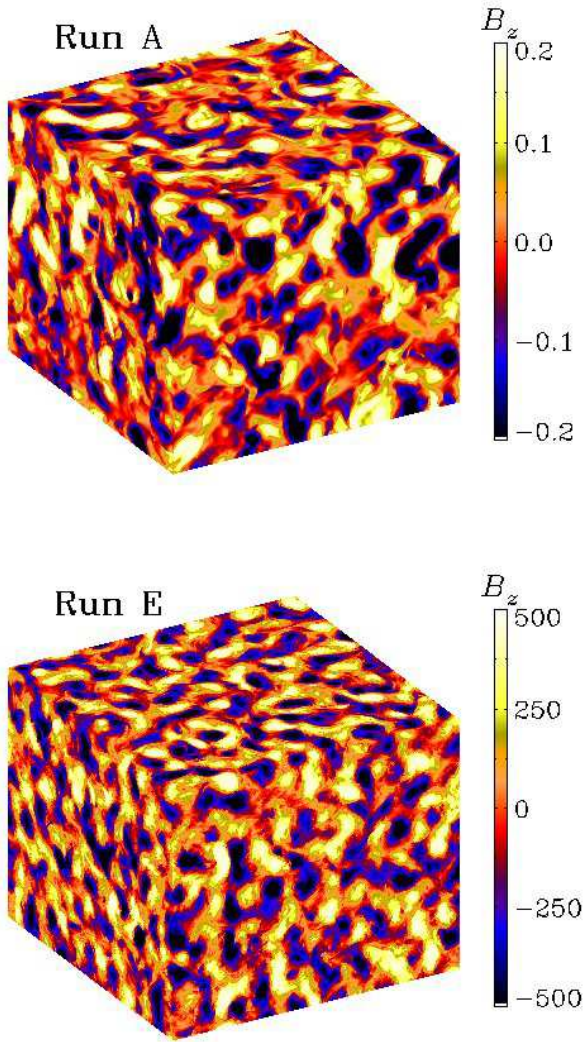}
\end{center}\caption[]{
Visualization of $B_z$ on the periphery of the
computational domain for Runs~A and E at a time
when $k_0\xi(t)\approx24$.
}\label{AE}\end{figure}

In \Fig{pkt_few_Hf1_t2em5_k180a} we plot $E(k,t)$ at times separated by
a factor of $10^{1/3}\approx2.15$.
We also plot the normalized helicity, $kH(k,t)/2$.
We see that it quickly begins to evolve underneath a flat envelope.
Note that the amplitude of the spectrum is unchanged with time, so
the exponent $\beta$ in \Eq{EXIbeta} must be zero.
The same behavior is seen in a case where the initial spectrum is driven
by short-term forcing; see \Fig{pkt_few_Hf1_t2em5_k180a}(b).
In the $pq$ diagram, the solution displays a drift of the point
$(p,q)$ along the $\beta=0$ line toward the point $p=q=2/5$;
see \Fig{pq_Hf1_t2em5_k180a}.

In \Fig{AE} we show visualizations of $B_z$ on the periphery of the
computational domain for Runs~A and E.
Both figures look remarkably similar, so the presence of helicity bears no
obvious imprint on such a scalar representation of the magnetic field.

\begin{figure}\begin{center}
\includegraphics[width=\columnwidth]{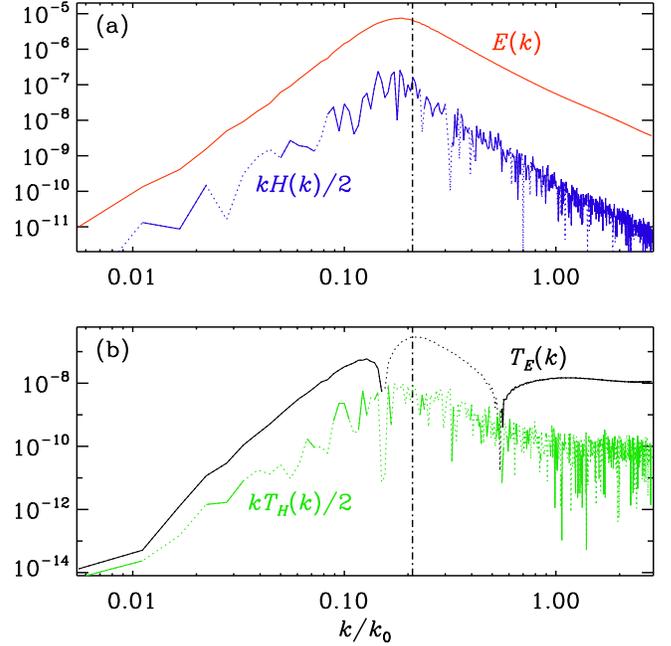}
\end{center}\caption[]{
(a) Magnetic energy spectrum (red) and scaled magnetic helicity spectrum
(blue), and (b) the corresponding transfer spectra of magnetic energy and
scaled magnetic helicity for Run~A.
Solid (dotted) line sections denote positive (negative) values.
}\label{pkt2_one_Hf0_t2em5_k180c}\end{figure}

\begin{figure}\begin{center}
\includegraphics[width=\columnwidth]{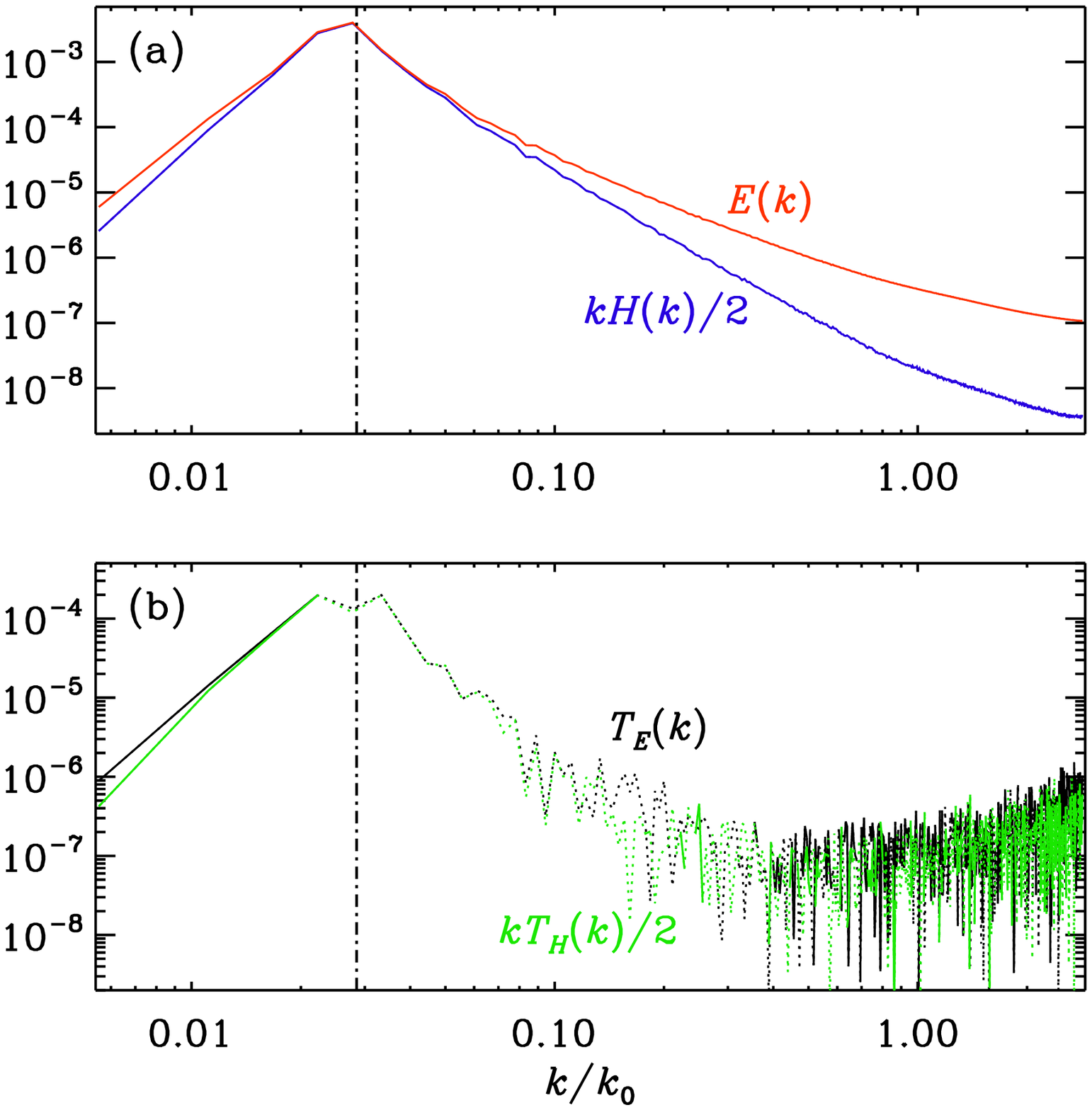}
\end{center}\caption[]{
Same as \Fig{pkt2_one_Hf0_t2em5_k180c}, but for Run~E.
}\label{pkt2_one_Hf1em3_t1em4_k180a}\end{figure}

In \Figs{pkt2_one_Hf0_t2em5_k180c}{pkt2_one_Hf1em3_t1em4_k180a}, we show
magnetic energy and helicity spectra along with the corresponding transfer
spectra at the last time in the simulation.
While in the helical case (Run~E), $kH(k,t)/2$ is almost equal to $E(k,t)$
for $k$ values near the position of the spectral magnetic energy peak,
it is about 10 times weaker in the nonhelical case (Run~A).
If time were to continue to grow, the difference between the two lines
would decrease further; cf.\ \Fig{pkt_few_Hf0_t2em5_k180a}.
The magnetic energy has a clear $k^5$ subinertial range spectrum and a
$k^{-3}$ inertial range spectrum.
The helicity spectra are a bit steeper both for $k\xi<1$ and $k\xi>1$.

\begin{figure}\begin{center}
\includegraphics[width=\columnwidth]{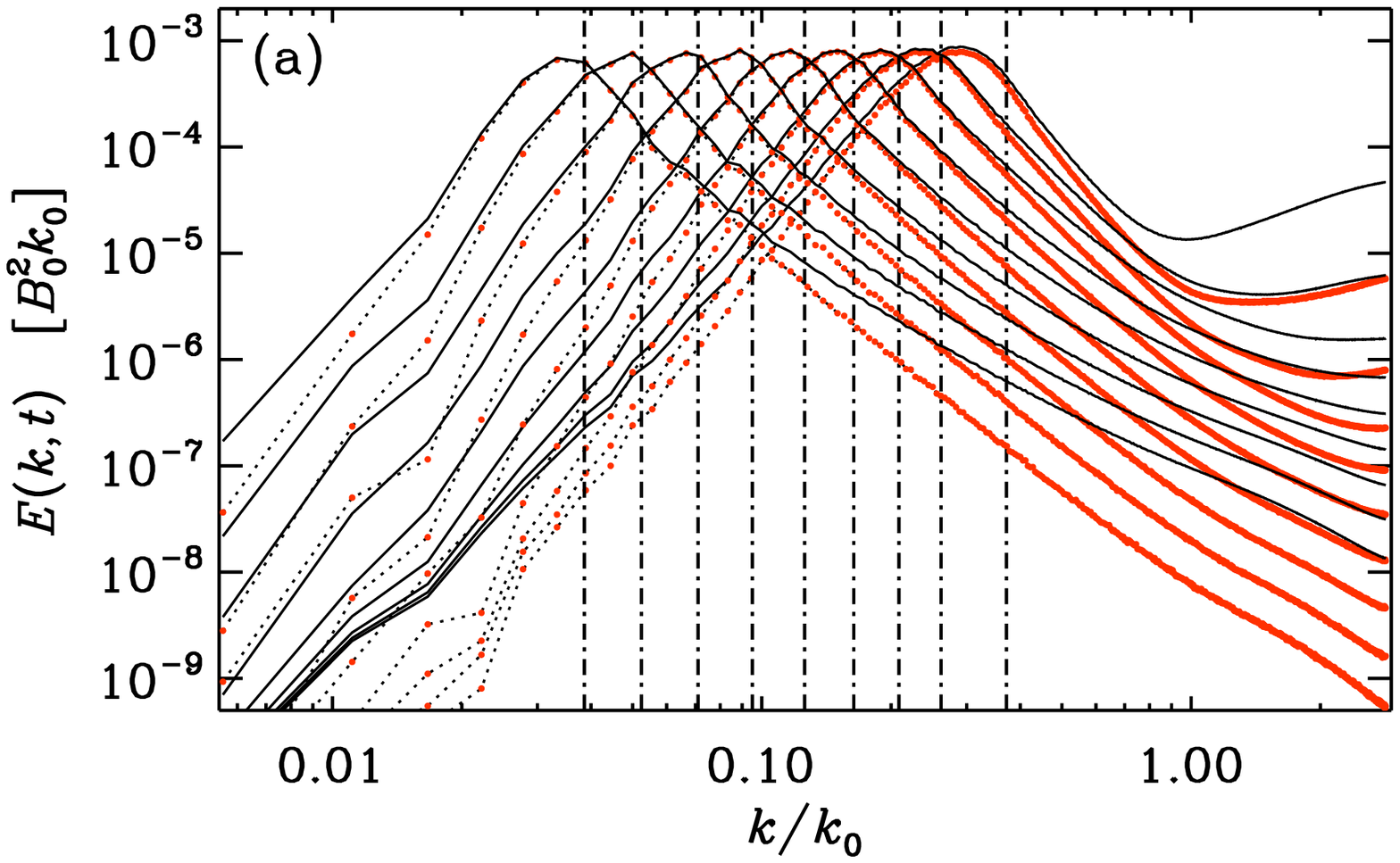}
\includegraphics[width=\columnwidth]{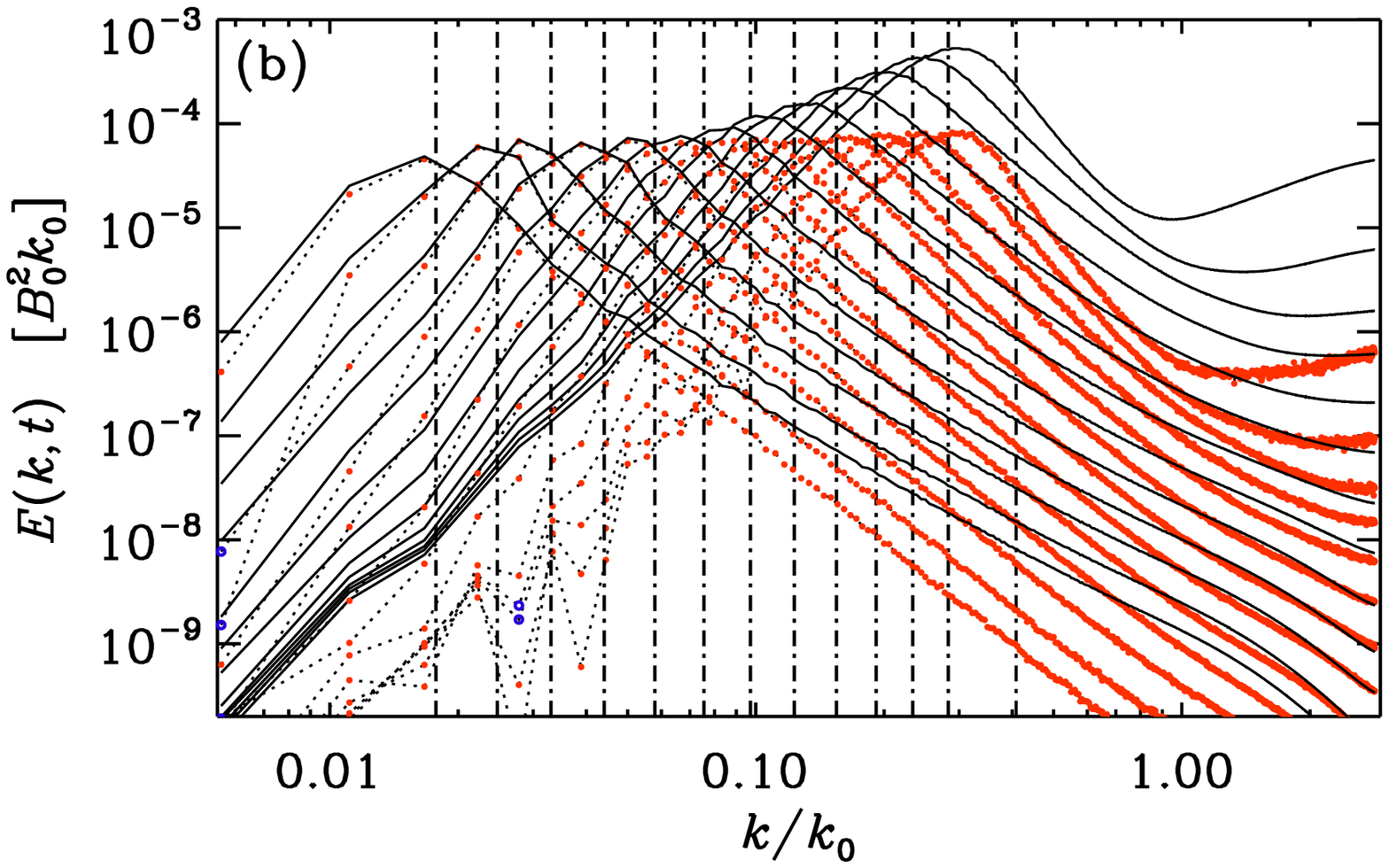}
\end{center}\caption[]{
Similarly to \Fig{pkt_few_f512c}, but for Runs~D and C and
fractional initial magnetic helicity with (a) $\sigma=10^{-2}$,
and (b) $\sigma=10^{-3}$.
}\label{pkt_few_Hf1em3_t2em5_k180a}\end{figure}

\subsection{Fractional magnetic helicity}

The case of fractional magnetic helicity is arguably the most important
case, because there is always some helicity in each hemisphere of any
rotating stratified body, and it is usually never hundred percent.
The ratio ${\cal H}/2\xi(t){\cal E}$ is between $-1$ and $+1$ and
is a measure of the degree of fractional magnetic helicity.
It turns out that, unlike the case of usual MHD \citep{TKBK12}, in the
Hall cascade, a very small amount of magnetic helicity ($\sigma=10^{-3}$)
can lead to nearly 100\% helicity in a moderate amount of time; see
\Fig{pkt_few_Hf1em3_t2em5_k180a}.

The consequences of the realizability condition become apparent when
comparing energy and helicity spectra at subsequent times in the same
plot; see \Fig{pkt_few_Hf1em3_t2em5_k180a}.
We see that the normalized helicity spectrum quickly begins to evolve
underneath a flat envelope (with no or a very small slope).

\begin{table}\caption{
Examples illustrating the realizability condition for Runs~A--C and E
at selected times.
}\vspace{12pt}\begin{tabular}{cccccccc}
Run & $t\eta_0k_0^2$ & ${\cal E}/{\cal E}_0$ & $k_0\xi$ &
$2\xi{\cal E}/{\cal H}_0$ & $\tilde{{\cal H}}/{\cal H}_0$ &
$|{\cal H}|/{\cal H}_0$ \\
\hline 
A &   0.4  &  1.0000 &   1.9 & 298.49 &  1.30 &  1.00 \\
  &   1.7  &  0.3537 &   2.8 & 157.51 &  0.96 &  0.78 \\
  &  17    &  0.0513 &   4.8 &  39.24 &  0.54 &  0.52 \\
  & 170    &  0.0062 &   8.6 &   8.59 &  0.33 &  0.32 \\
  &1700    &  0.0007 &  16.0 &   1.89 &  0.18 &  0.18 \\
  &8100    &  0.0002 &  24.1 &   0.70 &  0.11 &  0.11 \\
\hline 
B &   0.17 &  1.000 &   2.5 & 150.82 &  1.17 &  1.00 \\
  &   1.7  &  0.192 &   5.2 &  61.19 &  0.95 &  0.90 \\
  &  17    &  0.032 &   9.9 &  19.70 &  0.82 &  0.80 \\
  & 170    &  0.005 &  19.4 &   6.54 &  0.70 &  0.70 \\
  & 810    &  0.002 &  30.7 &   3.08 &  0.65 &  0.65 \\
\hline 
C &   0.17 &  1.0000 &   2.4 &   7.95 &  1.00 &  1.00 \\
  &   1.7  &  0.1956 &   5.2 &   3.31 &  0.90 &  0.90 \\
  &  17    &  0.0378 &  10.3 &   1.26 &  0.81 &  0.81 \\
  & 170    &  0.0107 &  22.7 &   0.79 &  0.73 &  0.73 \\
  &1700    &  0.0041 &  56.4 &   0.76 &  0.68 &  0.68 \\
\hline 
E &   0.0017& 1.0000 &   1.1 &   1.06 &  1.00 &  1.00 \\
  &   0.017&  0.5207 &   2.2 &   1.08 &  1.02 &  1.02 \\
  &   0.17 &  0.1981 &   5.7 &   1.05 &  1.01 &  1.01 \\
  &   1.7  &  0.0785 &  14.1 &   1.04 &  0.98 &  0.98 \\
  &  17    &  0.0313 &  35.1 &   1.03 &  0.97 &  0.97 \\
\label{Thel}\end{tabular}
\tablenotemark{
${\cal E}_0$ and ${\cal H}_0$ denote the initial values of
${\cal E}$ and ${\cal H}$, respectively.
}\end{table}

\begin{figure}\begin{center}
\includegraphics[width=\columnwidth]{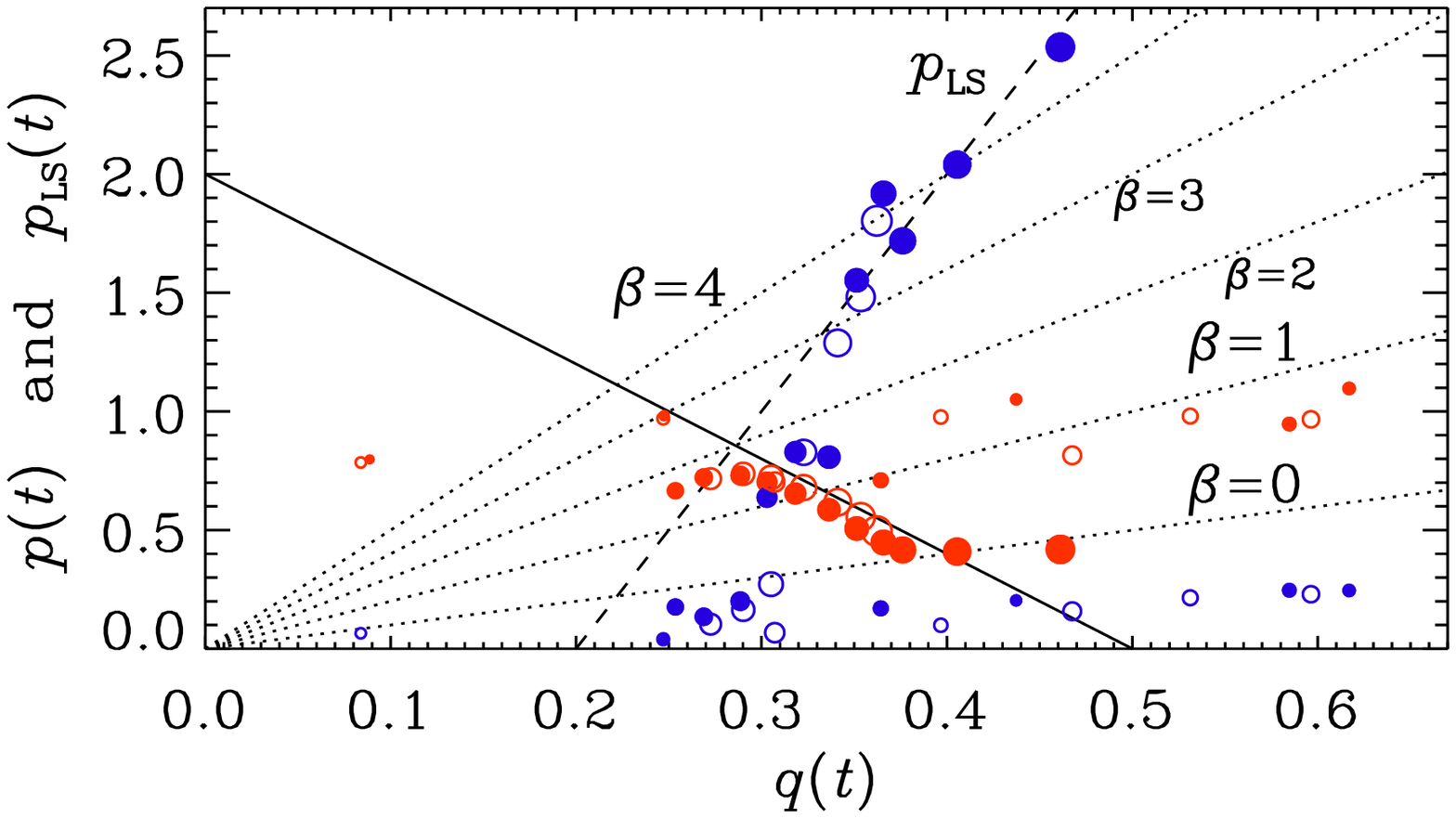}
\end{center}\caption[]{
Similar to \Fig{pq_Hf0_t2em5_k180c}, but for Runs~C (for filled symbols)
and C' (for open symbols).
}\label{pq_Hf1em3_t2em5_k180a}\end{figure}

The spectral magnetic energy is initially much larger than $kH(k,t)/2$,
because the fractional helicity is small.
Rather soon, however, the magnetic energy spectrum reaches $kH(k)/2$
and cannot drop any further.
By that time, the fractional magnetic helicity has reached nearly
hundred percent at $k\approx\xi^{-1}$.
Both at smaller and larger $k$, however, the magnetic field cannot
reach hundred percent, presumably because of a direct cascade of
current helicity for $k\xi\gg1$, as in ordinary MHD \citep{BS05b}.
A current helicity cascade makes $k^2H(k)$ nearly parallel to $E(k)$,
so $kH(k)/2$, which is what is plotted, falls off faster than $E(k)$.
Also, for $k\xi\ll1$, the $kH(k)/2$ spectrum is steeper than
that of $E(k)$, and it even changes sign.
Looking at \Fig{pkt2_one_Hf1em3_t1em4_k180a}, however, we see that the
difference between $kH(k)/2$ and $E(k)$ is much less for $k\xi\ll1$
than for $k\xi\gg1$, so the steeper slope of $kH(k)/2$ for $k\xi\ll1$
may not be significant.

To illustrate the effect of the realizability condition, we recall
that, dividing \Eq{RealSpec} by $k$ and integrating over $k$, we obtain
\EQ
2\xi{\cal E}\equiv2\int k^{-1}E(k,t)\,dk \ge \int |H(k,t)|\,dk \ge |{\cal H}|.
\EN
In \Tab{Thel} we list the four terms for Runs~A--C and E at some
selected times, where we have denoted the penultimate term by
$\tilde{{\cal H}}\equiv\int |H(k,t)|\,dk$.
For the nonhelical runs (A and B), we find $\tilde{\cal H}>|{\cal H}|$
at early times, but even Run~C with fractional magnetic helicity obeys
$\tilde{\cal H}=|{\cal H}|$ already at early times.
At late times, the fractional helicity, ${\cal H}/2\xi{\cal E}$, is
still only 21\% for Run~B and 16\% for Run~A.

The $pq$ diagram in \Fig{pq_Hf1em3_t2em5_k180a} confirms that the
point $(p,q)$ evolves along the selfsimilarity line from $\beta=2$
toward $\beta=0$.
There is actually still a small separation between the red filled symbols
and the self similarity line.
In this connection, we recall that we used in all of our runs the value
$r=0.43$, although for $\beta\to0$, the value $r=0.20$ would have been
more appropriate.
This is indeed the case; see the open symbols in
\Fig{pq_Hf1em3_t2em5_k180a}, for are done for Run~C'.

\subsection{Dissipation}

An important outcome of our models is the resulting Joule dissipation.
A certain fraction of this energy supply is believed to power
the observed X-ray emission of the central compact objects of supernova
remnants \citep{GWH16,GWH18,GHI20}.
However, there is also neutrino emission from the crust \citep{Vig+13}.
Therefore, there is no direct equivalence between Joule dissipation and
the final X-ray emission, which depends on a number of factors, including
the resulting thermal stratification of the crust; see \cite{PV19}
for a review.
With this caveat in mind, we must consider the ``total'' luminosities
quoted below as upper limits for the X-ray luminosity.

\begin{figure}\begin{center}
\includegraphics[width=\columnwidth]{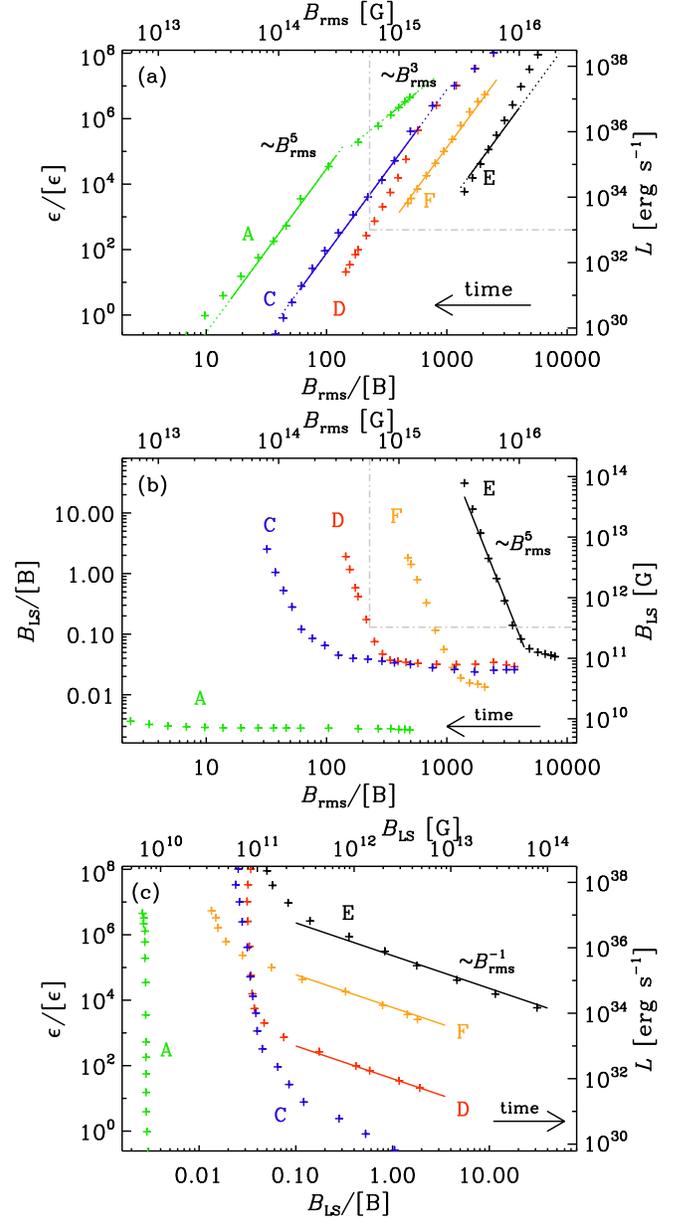}
\end{center}\caption[]{
(a) Magnetic dissipation and (b) large-scale magnetic field versus $\Brms$
for Runs~A (green), C (blue), D (red), E (orange), and F (black).
The plus signs denotes different times increasing from right to left,
separated by a factor $2.15$.
The straight lines in (a) and (b) show the $\Brms^5$ and $\Brms^{-5}$
scalings, respectively, with dotted sections denoting departures from
the plus symbols.
In (a), we also show the $\Brms^3$ scaling for the early phase of Run~A.
}\label{pdecay_comp}\end{figure}

We have computed the Joule dissipation $\epsilon$ from our models as a
function of time.
As $\Brms$ decreases, $\epsilon$ also decreases.
It is useful to plot $\epsilon$ versus $\Brms$.
The result is shown in \Fig{pdecay_comp} for Runs~A, and C--F.
In panel (a), we clearly see that $\epsilon$ is a very steep function
of $\Brms$.
At the same time as $\Brms$ decreases, the large-scale field increases.
This is shown in panel (b) where we plot
$B_{\rm LS}=(2{\cal E}_{\rm LS})^{1/2}$ versus $\Brms$.
These dependencies are also very steep---inversely proportional to be
$\Brms^5$.
Since $\epsilon\propto\Brms^5$ and $B_{\rm LS}\propto\Brms^{-5}$,
we have $\epsilon\propto B_{\rm LS}^{-1}$; see \FFig{pdecay_comp}(c).

Our work has shown that the magnetic energy and correlation length depend
in power law form on time with ${\cal E}\propto t^{-p}$ and $\xi\propto
t^q$, respectively, where $p=6/7$ and $q=2/7$ in the nonhelical case
with $\beta=2$ and $p=q=2/5$ in the helical case with $\beta=0$; see
\Tab{pq_tab}.
Since Joule dissipation is given by $\eta\mu_0\JJ^2$, we expect
$\epsilon\propto\eta{\cal E}\xi^{-2}$, and therefore
$\epsilon\propto t^{r-p-2q}\propto\Brms^s$, with $s=2(p+2q-r)/p$.
Thus, we have $s=7/3\approx2.33$ in the nonhelical case with $\beta=2$
and $s=5$ in the helical case with $p=q=2/5$.

Most of our results show a dependence compatible with
$\epsilon\propto\Brms^5$.
Run~A shows initially a shallower scaling close to $\Brms^3$
(which would be expected for $\beta=1$), although the theoretically
expected scaling would be shallower.
The scaling of the red line in \Fig{pdecay_comp} does perhaps best
describe the values proposed by \cite{GWH16,GHI20} for their global model
of NS crusts with $\ell=10$.
The gray lines in \Fig{pdecay_comp} highlight a particular example
with a total luminosity $L_{\rm tot}=10^{33}\erg\s^{-1}$,
which corresponds to $\Brms\approx6\times10^{14}\G$.
Thus, we can write
\EQ
L_{\rm tot}=10^{33}\,\left(\frac{\Brms}{6\times10^{14}\G}\right)^5\erg\s^{-1}
\quad\mbox{(red line)}.
\EN
\FFig{pdecay_comp}(b) shows that then $B_{\rm LS}\approx3\times10^{11}\G$.

We also see from \Fig{pdecay_comp}(b) that nonhelical magnetic fields do
not produce significant large-scale magnetic fields (see the green curve).
However, because those fields decay much more rapidly, they also dissipate
more energy for a given field strength.
If such a scenario was to be viable, one would need to have a preexisting
large-scale magnetic field.
This could lead to more rapid magnetic field decay \citep{Bra+20} and would
need to be studied more carefully.

As already emphasized by \cite{GHI20}, only the large-scale magnetic field
of NSs can be observationally inferred from the spin-down of pulsars.
Therefore, we show in \Fig{pdecay_comp}(c) $\epsilon$ versus $B_{\rm LS}$.
We see that $\epsilon$ now decreases with increasing $B_{\rm LS}$.
The slope is $-1$ for fractionally helical magnetic fields at late times,
as expected from the aforementioned quintic scalings of $\epsilon$
and $B_{\rm LS}$ with $\Brms$.

\begin{table}\caption{
Parameters for the stratified models.
}\vspace{12pt}\begin{tabular}{ccccc}
Run & $\sigma_0$ & $\phi_0(k_1z=-6)$ & $\phi_0(k_1z=-3)$ & $\phi_0(k_1z=-1)$ \\
\hline
Bz &   0   & 0.0007 &  ---   & 0.0070 \\
Cz & 0.001 & 0.006  & 0.012  & 0.043  \\
Dz & 0.01  & 0.06   & 0.12   & 0.30   \\
Ez &   1   &  ---   &  ---   &  ---   \\
CZ & 0.001 & 0.0003 & 0.0033 & 0.013  \\
\hline
z  &$\quad\xi_0\rightarrow\!\!\!\!\!$
           & 0.003  & 0.005  & 0.009  \\
Z  &$\quad\xi_0\rightarrow\!\!\!\!\!$
           & 0.001  & 0.002  & 0.005  \\
\label{zTmodels}\end{tabular}
\tablenotemark{
Dashes indicate that a fit would be uncertain.
The resolution is $1024^3$.
}\end{table}

\begin{figure}\begin{center}
\includegraphics[width=\columnwidth]{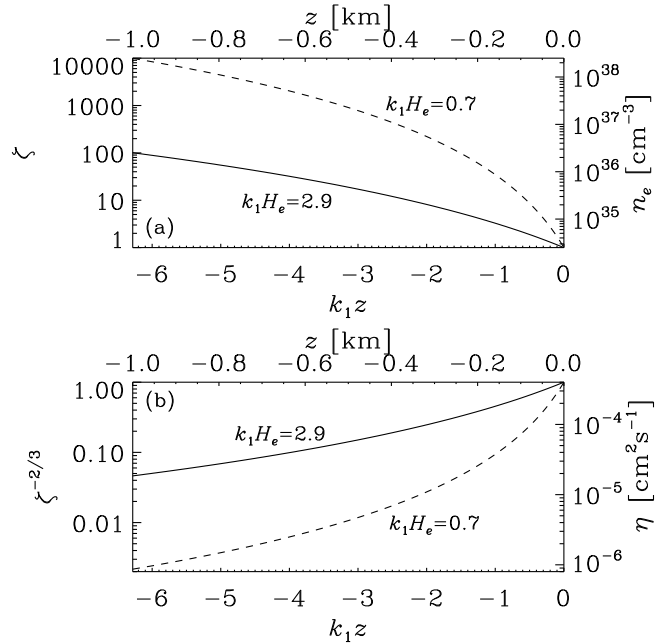}
\end{center}\caption[]{
(a) $\zeta$ (left axis) and $n_e$ (right axis), and
(b) $\zeta^{-2/3}$ (left axis) and $\eta$ (right axis)
versus $k_1z$ (lower axes) and $z$ (upper axes)
for $k_1H_e=2.9$ (solid lines) and $k_1H_e=0.7$ (dashed lines).
}\label{pstrat}\end{figure}

\subsection{Dissipation in the stratified case}

Real NS crusts are strongly stratified with $n_e$ and $\eta$ varying
over several orders of magnitude.
The radial variation of $n_e$ was already considered by \cite{VCO00} and
\cite{HR04}, but they assumed $\eta$ to be constant.
Approximately realistic profiles for both $n_e$ and $\eta$ were adopted
in the works of \cite{PG07} and \cite{Vig+13} in two dimensions and
\cite{GWH16,GHI20} in three dimensions.
Here we adopt the prescription of \cite{GWH16,GHI20}, who used
the value $H_e/R=0.0463$ for the scale height.
With $d/R=0.1$, this corresponds to $k_1 H_e=(2\pi/d)\,H_e\approx2.9$.

The resulting profiles of $n_e\propto\zeta(z)$ and
$\eta\propto\zeta^{-2/3}$ are shown in \Fig{pstrat}.
For comparison with a stronger and even more realistic
stratification, we also include a case with $k_1H_e=0.7$.
We see that at the surface, we have the values
$n_e=2.5\times10^{34}\cm^{-3}$ and $\eta=4\times10^{-4}\cm^2\s^{-1}$
that were quoted in \Sec{UnitsNS} in SI units.
These values were also used by \cite{GWH16,GHI20}, where the surface
conductivity was $\sigma_{\rm el}=1.8\times10^{23}\s^{-1}$.
This value is similar to that of \cite{PG07}, who give surface values for
$\sigma_{\rm el}$ of around $10^{23}\s^{-1}$ for temperatures of slightly
below $10^8\K$, and somewhat larger values for smaller temperatures.
This conductivity corresponds to
$\eta=c^2/4\pi\sigma_{\rm el}=7\times10^{-4}\cm^2\s^{-1}$ in cgs units,
where $c=3\times10^{10}\cm\s^{-1}$ is the speed of light.
In the strongly stratified case with $k_1H_e=0.7$, $n_e$ and $\eta$
vary between top and bottom of the domain by four and nearly three orders
of magnitude, respectively.
The latter variation is similar to that in Figure~1 of \cite{PG07}.

We then solve \Eq{dAdt2} with nonperiodic boundary conditions in
the $z$ direction.
We have calculated models for different values of the initial helicity
parameter $\sigma_0$; see \Tab{zTmodels} for Runs~Bz--Ez, which have the
same values of $\sigma_0$ as the unstratified counterparts, Runs~B--E.
We also list Run~CZ, which has $k_1H_e=0.7$ and $\sigma=0.001$
(like Runs~C and Cz).
As before, we use $k_0/k_1=180$ in all stratified cases.

\begin{figure}\begin{center}
\includegraphics[width=\columnwidth]{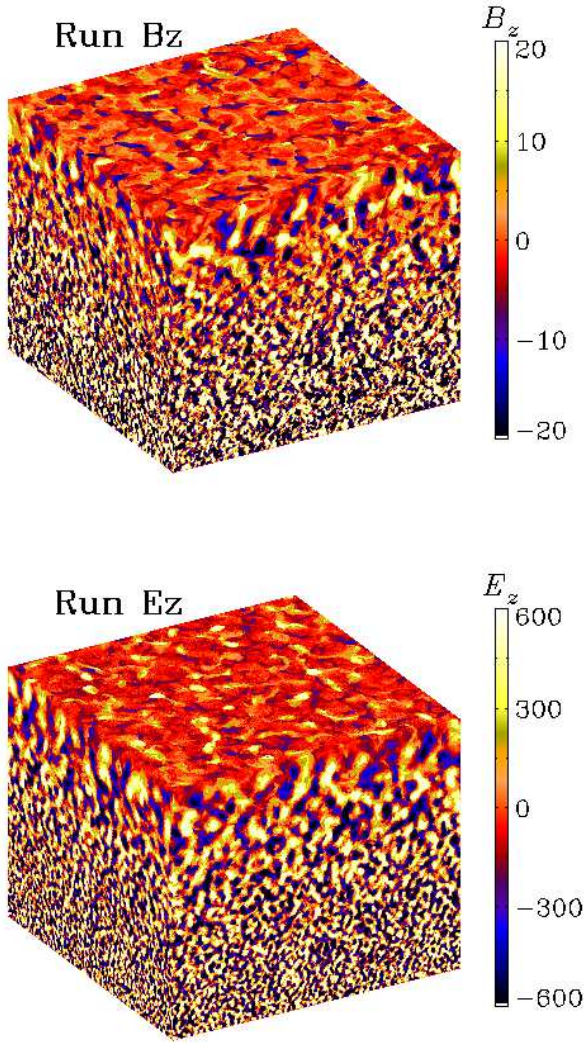}
\end{center}\caption[]{
Similar to \Fig{AE}, but for Runs~Bz and Ez.
}\label{BzEz}\end{figure}

In \Fig{BzEz} we show $B_z$ for Runs~Bz and Ez at times
$\eta_0 k_0^2 t=80$ and $3.5$, respectively, which is when
the correlation lengths are comparable in the two runs.
We see that the magnetic structures have larger length scales in the
upper layers.
This is primarily a consequence of the larger magnetic diffusivity there.
In the fully helical case (Run~Ez), the formation of larger length scales
is accelerated by the presence of magnetic helicity.

\begin{figure}\begin{center}
\includegraphics[width=\columnwidth]{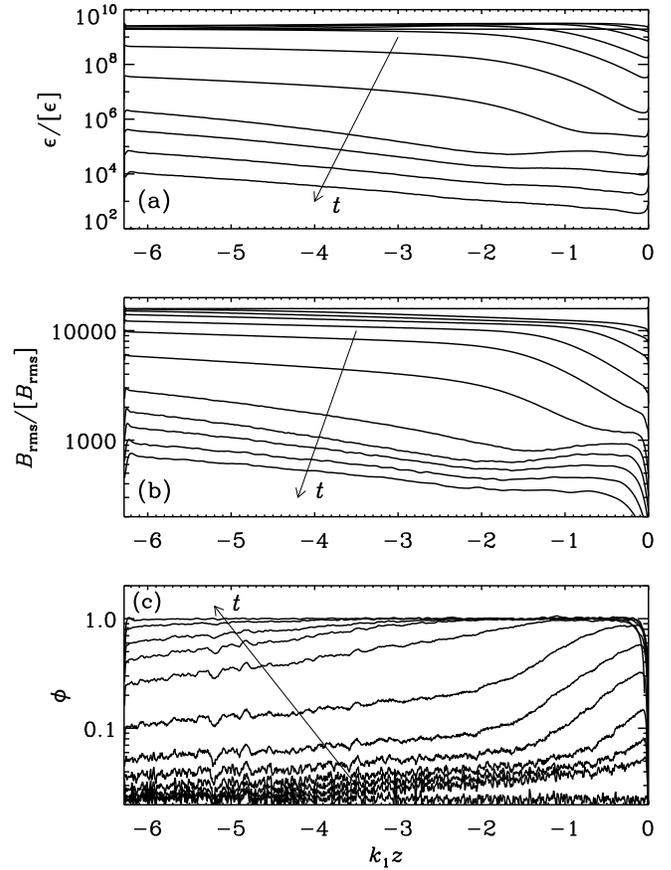}
\end{center}\caption[]{
(a) Dissipation, (b) rms magnetic field, and (c) fractional
magnetic helicity versus height for Run~Dz for increasing times,
as indicated by the arrows, in steps of $10^{1/2}\approx3.2$
until $\eta_0 k_0^2 t=140$.
}\label{ppxyaver}\end{figure}

To determine the fractional magnetic helicity as a function of $z$ and
$t$, we now employ a method that does not require Fourier transformation.
The relevant information is contained in the ratio,
\EQ
\overline{\AAA\cdot\BB}/\overline{\BB^2}\equiv\phi\xi,
\EN
where $\phi(z,t)$ is the fractional magnetic helicity, $\xi(z,t)$ is a
suitability defined correlation length, and overbars denote $xy$ averaging.
Instead of using \Eq{xi_def}, which requires Fourier transformation,
we compute $\xi$ from the ratio
\EQ
\xi^2=\overline{\AAA\cdot\BB}/\overline{\JJ\cdot\BB}.
\EN
It is then convenient to determine $\phi$ from the quantity
\EQ
\phi^2=\overline{\AAA\cdot\BB}\;\overline{\JJ\cdot\BB}/(\overline{\BB^2})^2.
\EN
For very small fractional helicities, however, $\phi^2$ can occasionally
become negative in some places.
In all other cases, however, it is possible to compute $\phi(z,t)$.

In \Fig{ppxyaver} we plot the $z$ dependence of $\epsilon$, $\Brms$,
and $\phi$ at different times.
It turns out that $\epsilon$ is largest in the deeper parts.
This is mainly a consequence of $\Brms$ having decayed most rapidly
near the surface, where $\eta$ is large.
We also see from \Fig{ppxyaver}(c) that $\phi$ grows fastest in the
upper layers.

\begin{figure}\begin{center}
\includegraphics[width=\columnwidth]{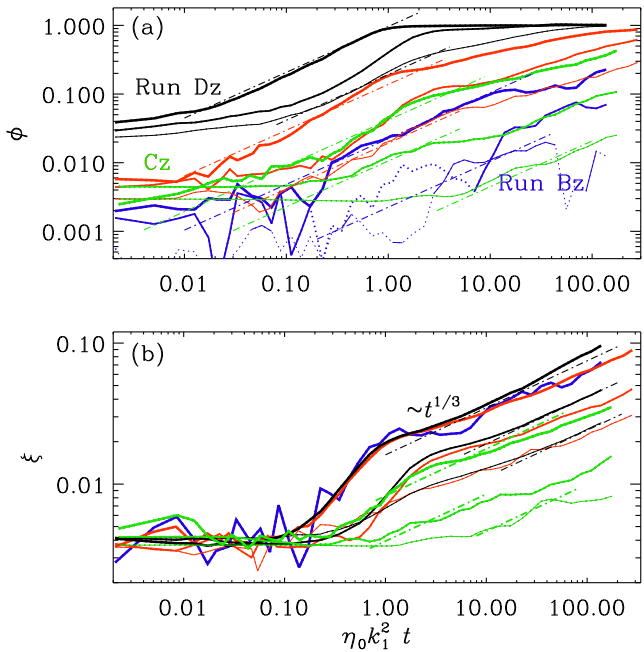}
\end{center}\caption[]{
(a) $\phi(t)$ for Runs~Bz (blue), Cz (red), Dz (back), and CZ (green) at
$k_1 z=-6$, $-3$, and $-1$, as indicated by the increasing line thickness.
Dotted lines fragments indicate $|\phi|$ when $\phi$ is negative.
The dash-dotted lines indicate the fits with the $\phi_0(z)$ given
in \Tab{zTmodels}.
(b) $\xi(t)$ for $k_1 z=-1$ for Runs~Bz--Dz and CZ.
}\label{ppxyaver_comp}\end{figure}

To study the growth of $\phi$ in more detail, we compare in
\Fig{ppxyaver_comp} the time dependence for Runs~Bz--Dz
and CZ for three values of $z$.
It turns out that, in an intermediate range, $\phi$ grows approximately
algebraically like
\EQ
\phi(z,t)=\phi_0(z)\,(t/t_0)^{2/3}.
\EN
where $t_0$ was defined in \Eq{etat_formula}.
The values of $\phi_0(z)$ are listed in \Tab{zTmodels}
for $k_1z=-6$, $-3$, and $-1$.
The magnetic helicity production in the initially nonhelical Run~B is
a result of random fluctuations and could equally well have been of
the opposite sign.
In that case, $\phi^2$ would normally still be positive, because the sign
of magnetic helicity affects the signs of both $\overline{\AAA\cdot\BB}$
and $\overline{\JJ\cdot\BB}$.
The square root of their ratio gives $\xi$ and is shown in the
lower panel of \Fig{ppxyaver_comp}.
It also obeys power law scaling of the form
\EQ
\xi(z,t)=\xi_0(z)\,(t/t_0)^{1/3},
\EN
where the $\xi_0(z)$ are similar for all three runs, suggesting that
their values are approximately independent of the magnetic helicity
for fixed stratification.
The $\xi_0(z)$ increase with $z$; see the last two lines of \Tab{zTmodels}
for models of series z and Z with $k_1H_e=2.9$ and $0.7$, respectively.

\begin{figure}\begin{center}
\includegraphics[width=\columnwidth]{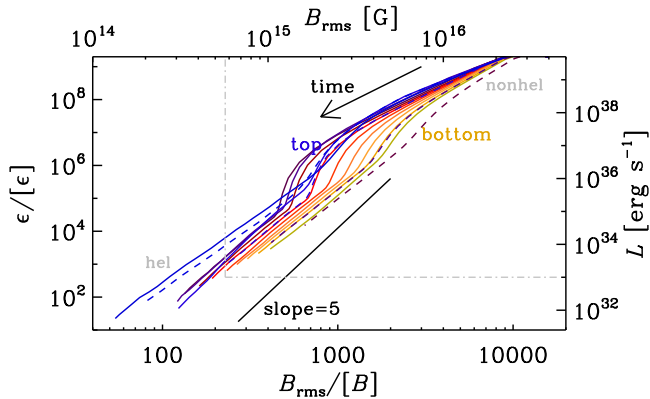}
\end{center}\caption[]{
Similar to \Fig{pdecay_comp}(a), but for Run~Cz (solid lines) and Run~CZ
(dashed lines), showing the dependencies separately for each horizontal
layer, except for Run~CZ, where only the surface layers with comparable
values of $\zeta$ are plotted.
Yellow (blue) shades indicate locations near the bottom (top)
of the domain.
}\label{ppxyaver_parametric}\end{figure}

An important question is how the dissipation properties in the cases with
stratification compare with the unstratified cases.
We therefore show again a parametric representation of $\epsilon$
versus $\Brms$, but now for each horizontal layer separately.
The result is shown in \Fig{ppxyaver_parametric}.
Interestingly, the different curves tend to collapse on top of each other.
Furthermore, we obtain a similar $\epsilon\propto\Brms^5$ scaling as
in the unstratified case, but it is now along two branches that are
slightly offset relative to each other.
The upper branch is somewhat shallower ($\epsilon\propto\Brms^3$).
Nevertheless, the values of $\epsilon$ and $\Brms$ are remarkably
similar to those in the unstratified cases, and are again compatible with
$L\approx10^{33}\erg\s^{-1}$ for a $\Brms$ of a few times $10^{14}\G$.
This can be seen by comparing with the gray dashed-dotted
lines in \Fig{pdecay_comp}, which have also been reproduced in
\Fig{ppxyaver_parametric} at the same position.

The relatively good agreement between \Fig{pdecay_comp}(a) and
\Fig{ppxyaver_parametric} provides some justification for using the
unstratified models as a meaningful local representation of the Hall
cascade in NS crusts.
The reason why unstratified models were previously found to yield a poor
representation of the case with stratification \citep{PG10} is probably
related to the fact that those authors studied a large-scale nonturbulent
magnetic field, which is more sensitive to boundary conditions.

The sudden drop in $\epsilon$ for intermediate $\Brms$
is probably caused by the gradual transition from a fractionally helical
to a fully helical magnetic field, which occurs here not only as a function of
time, but also as a function of $z$.
Near the surface, as already shown in \Fig{ppxyaver_comp}, this transition
happens earlier than in the deeper parts.

The work of \cite{GWH16,GHI20} only considered the case of moderately
strong stratification, where $k_1H_e=2.9$.
In addition to possible numerical problems, there is always the general
difficulty for stronger stratification (smaller values of $k_1H_e$)
that the characteristic time scales in the problem become
very different between top and bottom of the domain.
The time step is limited mainly by the Hall nonlinearity in the surface
layers, but the evolution in the deeper layers becomes very slow, so we
need a large number of time steps to describe the full time evolution.

\section{Conclusions}

The present work has confirmed that the Hall cascade can liberate
a significant amount of dissipative energy through Joule heating.
The resulting heating is proportional to $\Brms^5$ for helical and
fractionally helical magnetic fields.
The magnetic fields undergo strong inverse cascading with a temporal decay
significantly slower ($\propto t^{-2/5}$) than in MHD ($\propto t^{-2/3}$).
However, even in the nonhelical cases, there can be inverse cascading,
but only for strong magnetic fields.

We confirmed the $k^{-7/3}$ inertial range spectrum both in forced and
decaying cases.
The nondimensional coefficient $C_{\rm Hall}$ in this relation has
been determined to be approximately $1.6$ in the helical case and
$2.7$ in the nonhelical case.
However, this was not a major focus of attention and more accurate
determinations should be performed using dedicated higher resolution
simulations.
The current helicity cascade, expected to be proportional to $k^{-2}$,
is also worth reconsidering.
Furthermore, in the decaying case, we find a steeper subinertial range
spectrum proportional to $k^5$.
It develops independently of the subinertial range slope of the initial
field.

Most of our models predict a rather sensitive dependence of the heating
rate on the magnetic field strength proportional to the fifth power of
the rms value.
It would be useful to confirm the generality of this scaling using global
models such as those used by \cite{GWH16,GWH18,GHI20}.
These steep dependencies can potentially be employed as sensitive
diagnostic tools that may give us information about the dominant
physical processes leading to the X-ray emission for the central
compact objects of supernova remnants of different ages.
Before doing this, however, it would be necessary to establish
the detailed connection between Joule dissipation and X-ray luminosity
and to determine the contribution from neutrino emission.

The inverse cascade in the helical case has been seen before
\citep{Cho11}, but it was analyzed only at a qualitative level.
We have quantified this here by plotting the exponent $p_{\rm LS}$ as
a function of $q$, which we find to be compatible with a linear relation.
For the helical case, we find $p_{\rm LS}\approx2$, which implies
an approximately quadratic growth of the large-scale field with time.
The existence of such a relation was not anticipated.
However, our phenomenological scaling relation tends to predict slightly
smaller values of $p_{\rm LS}$ in some cases.
This could be related to finite size effects that lead to a slightly
shallower spectrum at small $k$ and thereby to larger ${\cal E}_{\rm LS}$,
which could explain the faster growth.
How significant this departure is remains unclear, so this too would be
worth reconsidering.

The mechanism for inverse cascading in the nonhelical case is not very
clear either.
It is possible that magnetic helicity fluctuations could be responsible
for this.
The usual fractional magnetic helicity is clearly too small, but even
the value based on the modulus of the magnetic helicity spectrum,
$\tilde{{\cal H}}=\int |H(k,t)|\,dk$, is small and is equal to
$|{\cal H}|$ at late times; see \Tab{Thel}.
Therefore, this explanation might not be fully satisfactory.

At late times, our simulations display a self-similar decay.
We have seen that the correlation length $\xi$ increases by a
factor of around a hundred by the end of our simulations; see
\Fig{pkt_few_Hf1em3_t2em5_k180a}(b).
For later times, one would need to allow for the finite extent of the
global spherical shell geometry of NS crusts, as was already done by
\cite{GWH16,GWH18,GHI20}.
They considered the peak of the spectral magnetic energy to be at a
spherical harmonic degree of around $\ell=10$--$20$, which corresponds
to ``effective'' values of $k_0$ that are already comparable to $k_1$,
the smallest vertical wavenumber in our domain.
We must therefore regard our local simulations as the early stages of
a selfsimilar evolution, after which the finite shell geometry is best
described by global models with $\ell\sim R/\xi$.

Most of our attention went into the study of local unstratified
models with periodic boundary conditions.
However, it turns out that much of the physics of the unstratified models
can be recovered in the stratified ones with nonperiodic boundaries at
the appropriate depth.
It will be interesting to see whether this similarity between stratified
and unstratified models persists also when studying models where
the local temperature evolution is taken into account.
Technically, this should well be feasible with the {\sc Pencil Code}.
In this connection, we remind the reader that both the code and the
input files are freely available.
It would also be useful to couple the present studies to models of the
very early phases of NSs.
One would then be able to relax the assumption of an initial magnetic
field that was here assumed to have the same rms value at all heights.
It should be emphasized, however, that the earlier onset of growth of
magnetic helicity in the upper layers of our models is not connected with
our choice of the initial condition and is just a consequence of $\eta$
being larger near the surface.

\section*{Acknowledgements}

I thank the referee for the suggestion to compare with a stratified case.
This work was supported in part through the Swedish Research Council,
grant 2019-04234, and the National Science Foundation under the grant
AAG-1615100.
We acknowledge the allocation of computing resources provided by the
Swedish National Allocations Committee at the Center for Parallel
Computers at the Royal Institute of Technology in Stockholm.

\vspace{2mm}\noindent
{\large\em Software and Data Availability.} The source code used for
the simulations of this study, the {\sc Pencil Code} \citep{PC},
is freely available on \url{https://github.com/pencil-code/}.
The DOI of the code is https://doi.org/10.5281/zenodo.2315093.
The simulation setup and the corresponding data are freely available on
\url{https://doi.org/10.5281/zenodo.3951873}.


\end{document}